\documentclass[10pt]{amsart}
\usepackage{amsbsy,amssymb,amscd,amsfonts,latexsym,amstext,delarray,
amsmath,amsthm,graphicx}

\usepackage[all]{xy}

\newtheorem{thm}{Theorem}[section]
\newtheorem{prop}[thm]{Proposition}

\newtheorem{lem}[thm]{Lemma}

\newtheorem{defn}[thm]{Definition}
\newtheorem{rem}[thm]{Remark}
\newtheorem{ex}[thm]{Example}

\numberwithin{equation}{section}

\def\bG{{\mathbb G}}

\def\bL{{\mathbb L}}

\def\bT{{\mathbb T}}
\def\bU{{\mathbb U}}

\def\A{{\mathbb A}}
\def\C{{\mathbb C}}

\renewcommand{\P}{{\mathbb P}}
\def\Q{{\mathbb Q}}
\def\Z{{\mathbb Z}}
\def\R{{\mathbb R}}

\newcommand{\one}{1\hskip-3.5pt1}

\def\fL{{\mathfrak L}}

\def\fT{{\mathfrak T}}

\def\cE{{\mathcal E}}
\def\cF{{\mathcal F}}

\def\cH{{\mathcal H}}

\def\cK{{\mathcal K}}
\def\cL{{\mathcal L}}
\def\cM{{\mathcal M}}

\def\cO{{\mathcal O}}
\def\cP{{\mathcal P}}

\def\cR{{\mathcal R}}

\def\cV{{\mathcal V}}

\newcommand{\ie}{{\it i.e.\/}\ }

\newcommand{\cf}{{\it cf.\/}\ }
\def\text{\hbox}

\def\Aut{{\rm Aut}}

\def\Hom{{\rm Hom}}

\newcommand{\csm}{{c_*}}
\newcommand{\jXY}{{J(X,Y)}}

\title{Algebro-geometric Feynman rules}
\author[Aluffi]{Paolo Aluffi}
\author[Marcolli]{Matilde Marcolli}
\address{Department of Mathematics \\ Caltech \\ Pasadena, CA 91125, USA}
\email{aluffi\@@caltech.edu}
\email{matilde\@@caltech.edu}
\address{Department of Mathematics \\ Florida State University \\
Tallahassee, FL 32306, USA}
\email{aluffi\@@math.fsu.edu}
\email{marcolli\@@math.fsu.edu}
\address{Max--Planck Institut f\"ur Mathematik  \\
Vivatsgasse 7 \\
Bonn, D 53111, Germany}
\email{aluffi\@@mpim-bonn.mpg.de}
\email{marcolli\@@mpim-bonn.mpg.de}

\begin{document}

\begin{abstract}
We give a general procedure to construct algebro-geometric Feynman rules, that is,
characters of the Connes--Kreimer Hopf algebra of Feynman graphs that factor through
a Grothendieck ring of immersed conical varieties, via the class of the complement of the 
affine graph hypersurface. In particular, this maps to the usual Grothendieck ring of
varieties, defining motivic Feynman rules. We also construct an algebro-geometric Feynman
rule with values in a polynomial ring, which does not factor through the usual Grothendieck
ring, and which is defined in terms of characteristic classes of singular varieties. This invariant
recovers, as a special value, the Euler characteristic of the projective graph hypersurface complement.
The main result underlying the construction of this invariant is a formula for the characteristic
classes of the join of two projective varieties. 
We discuss the BPHZ renormalization procedure in this algebro-geometric context and
some motivic zeta functions arising from the partition functions associated to motivic
Feynman rules.
\end{abstract}

\maketitle

\section{Introduction}

In \cite{AlMa} we presented explicit computations of classes in the Grothendieck
ring of varieties, of Chern--Schwartz--MacPherson characteristic classes, and by specialization
Euler characteristics, for some particular classes of graph hypersurfaces. The latter are
singular projective hypersurfaces associated to the parametric formulation of Feynman
integrals in scalar quantum field theories and have recently been the object of extensive
investigation (see \cite{Blo}, \cite{Blo2}, \cite{BEK}, \cite{BK}, \cite{Mar}, \cite{MaRej}).

The purpose of the present paper is to answer a question posed to us by the referee
of \cite{AlMa}. We describe the problem here briefly, along with the necessary background. 
All this will be discussed in more details in the body of the paper.

For us a Feynman graph $\Gamma$ will be a finite graph whose set of edges 
consists of internal edges $E_{int}(\Gamma)$ and external edges $E_{ext}(\Gamma)$.
Whenever we focus on invariants that only involve internal edges,
we can assume that $\Gamma$ is just a graph in the ordinary sense. 

Consider a graph $\Gamma$ consisting of two components,
$\Gamma = \Gamma_1 \cup \Gamma_2$. To each component we can
associate a corresponding graph hypersurface $X_{\Gamma_1}\subset \P^{n_1-1}$
and $X_{\Gamma_2}\subset \P^{n_2-1}$, where $n_i =\# E_{int}(\Gamma_i)$ is the
number of internal edges of the Feynman graph $\Gamma_i$. The Feynman integral
$U(\Gamma_i,p_i)$, with assigned external momenta $p_i=(p_i)_e$ for 
$e\in E_{ext}(\Gamma_i)$, is computed in the parametric form as an integral over
a simplex $\sigma_{n_i}$ of an algebraic differential form defined on the hypersurface
complement $\P^{n_i-1}\smallsetminus X_{\Gamma_i}$. The multiplicative property
of the Feynman rules implies that, for a graph $\Gamma= \Gamma_1 \cup \Gamma_2$,
one correspondingly has $U(\Gamma,p)=U(\Gamma_1,p_1)U(\Gamma_2,p_2)$, so
that it is customary in quantum field theory to pass from the partition function whose
asymptotic series involves all graphs to the one that only
involves connected graphs. A further simplification of the combinatorics of graphs that
takes place in quantum field theory is obtained by passing to the 1PI effective action,
which only involves graphs that are 2-edge-connected (1-particle irreducible in the
physics terminology), \ie that cannot be disconnected by removal of a single edge.  

The Connes--Kreimer theory \cite{CoKr}, \cite{CoKr2} (see also
\cite{CoMa-book}) shows that the Feynman rules define a character of the Connes--Kreimer
Hopf algebra $\cH$ of Feynman graphs. Namely, the collection of dimensionally
regularized Feynman integrals $U(\Gamma,p)$ of all the 1PI graphs of a given scalar 
quantum field theory defines a homomorphism of unital commutative algebras 
$\phi \in \Hom(\cH, \cK)$, where $\cK$ is the field of germs of meromorphic functions 
at $z=0\in \C$. The coproduct in the Hopf algebra is then used in \cite{CoKr} to obtain 
a recursive formula for the Birkhoff factorization of loops in the pro-unipotent complex
Lie group $G(\C)=\Hom(\cH,\C)$. This provides the counterterms and the renormalized
values of all the Feynman integrals in the form of what is known in physics as the 
Bogolyubov recursion, or BPHZ renormalization procedure. 

In particular, any character of the Hopf algebra $\cH$ can be thought of as a possible 
assignment of Feynman rules for the given field theory, and the renormalization procedure 
can be applied to any such character as to the case of the Feynman integrals. 
In turn, the characters need not necessarily take values in the field $\cK$ of convergent 
Laurent series for the BPHZ renormalization procedure to make sense. 

In fact, it was shown in \cite{EFGK}  how the same Connes--Kreimer recursive formula
for the Birkhoff factorization of loops continues to work unchanged whenever 
the target of the Hopf algebra character is a Rota--Baxter algebra of weight $\lambda=-1$. 
In the Connes--Kreimer case, it is the operator of projection of a Laurent series onto its 
divergent part that is a Rota--Baxter operator and the Rota--Baxter identity is 
what is needed to show that, in the Birkhoff factorization 
$\phi =(\phi_- \circ S)\star \phi_+$, with $S$ the antipode
and $\star$ the product dual to the coproduct, the two terms $\phi_\pm$ 
are also algebra homomorphisms.

When working in the algebro-geometric world of the graph hypersurfaces $X_\Gamma$, 
one would like to have ``motivic Feynman rules", namely an assignment of an Euler characteristic 
$\chi_{new}$ (the class in the Grothendieck ring of varieties is a universal Euler characteristic) to 
the graph hypersurface complements $\P^{n-1}\smallsetminus X_\Gamma$ with the property that,
in the case of graphs 
$\Gamma$ consisting of several disjoint components $\Gamma_1\dots,\Gamma_k$,
one has
\begin{equation}\label{chiwant}
\chi_{new}(\P^{n-1}\smallsetminus X_\Gamma)= 
\prod_i \chi_{new}(\P^{n_i-1}\smallsetminus X_{\Gamma_i}),
\end{equation}
as in the case of the Feynman integrals $U(\Gamma,p)=\prod_i U(\Gamma_i,p_i)$.
Here the graph hypersurface $X_\Gamma$ associated to a graph $\Gamma$
is defined as the hypersurface in $\P^{n-1}$ ($n=\# E_{int}(\Gamma)$)
given by the vanishing of the polynomial
\[
\Psi_{\Gamma}(t_1,\ldots, t_n) = \sum_{T\subset \Gamma} \prod_{e\notin T} t_e,
\]
with the sum over spanning forests $T$ of $\Gamma$, and the product of the 
edge variables $t_e$ of the edges $e$ of $\Gamma$ that are not in the
forest $T$. If $\Gamma=\Gamma_1\cup \Gamma_2$ is a disjoint union,
then clearly
\begin{equation}\label{PsiGamma12}
\Psi_\Gamma(t_1,\ldots,t_n) =\Psi_{\Gamma_1}(t_1,\ldots,t_{n_1}) 
\Psi_{\Gamma_2}(t_{n_1+1},
\ldots, t_{n_1+n_2}).
\end{equation}

One can see that the usual Euler characteristic does not satisfy the desired property \eqref{chiwant}.
In fact, if $\Gamma$ is not a forest,  one can see that the hypersurface complement $\P^{n-1}\smallsetminus X_\Gamma$ is a $\bG_m$-bundle over the product $(\P^{n_1-1}\smallsetminus X_{\Gamma_1})\times (\P^{n_2-1}\smallsetminus X_{\Gamma_2})$, hence its Euler characteristic vanishes and the multiplicative property cannot be satisfied. 

The question the referee of \cite{AlMa} asked us is whether there
exists a natural modification $\chi_{new}$ of the usual Euler characteristic that 
restores the multiplicative property
\eqref{chiwant}, thus giving an interesting example of algebro--geometric Feynman rules. The main
result of the present paper is to show that indeed such modifications of the Euler characteristic
exist and they can be obtained from already well known natural enhancements of the Euler
characteristic in the context of algebraic geometry. In particular, we produce one such invariant
obtained using classes in the Grothendieck ring of varieties, and one obtained using 
the Chern--Schwartz--MacPherson characteristic class of singular algebraic varieties, \cite{MacPher} 
\cite{MHSchwartz}, and we show that both descend from a common invariant that lives
in a ``Grothendieck ring of immersed conical varieties".  

The first case, where one considers an invariant with values in the usual Grothendieck ring, 
has the advantage that it is {\em motivic}, so it indeed defines 
``motivic Feynman rules" as the referee suggested, and also it is in general easier to compute
explicitly, while the second case where the invariant is constructed in terms of characteristic classes is more difficult to compute, but it has the advantage that it takes values in a more manageable 
polynomial ring. 
We discuss the meaning of the BPHZ renormalization procedure in the Connes--Kreimer form
for some of these invariants, using suitable Rota--Baxter operators on polynomial algebras.

\smallskip

The paper is structured as follows. We recall briefly in \S \ref{Friqft} the properties of
Feynman integrals and Feynman rules in perturbative scalar quantum field theory, as
those serve as a model for our algebro-geometric definition.  In \S \ref{FRAGSec} we
introduce the notion of algebro-geometric Feynman rules, by requiring that the multiplicative
invariants associated to graphs depend on the data of the affine hypersurface complement,
up to linear changes of coordinates. We show in \S \ref{UnivSec} that there is a universal
algebro-geometric Feynman rule that takes values in a suitably defined Grothendieck
ring of immersed conical varieties, $\cF$. We show how the values behave under simple
operations on graphs, such as bisecting and edge, connecting graphs by a vertex or an edge, etc. 
The existence of this universal algebro-geometric Feynman rule is based on the 
multiplicative property of the affine hypersurface complements over disjoint unions of
graphs, which does not hold in the projective setting. We then show in \S \ref{MotFrules} 
that the universal Feynman rule maps to a motivic Feynman rule with values in the 
usual Grothendieck ring of varieties, by considering varieties up to isomorphism instead 
of the more restrictive linear changes of coordinates. We give an explicit relation between
the class of the affine hypersurface complement and the class of the 
projective hypersurface in the Grothendieck ring. As a consequence of the basic
properties of algebro-geometric Feynman rules, we show in Proposition \ref{SB1PI}
that the stable birational equivalence class of the projective graph hypersurface of a
non-1PI graph is equal to $1$. We also discuss how the parametric
formulation of Feynman integrals, in the case with nonzero mass and zero external momenta,
may fit in the setting of algebro-geometric Feynman rules with values in the algebra of
periods. The issue of the divergences of these integrals is further discussed in \S \ref{RenSec}.
In \S \ref{CharSec} we introduce a different algebro-geometric Feynman rule, that
is obtained by mapping the ring $\cF$ to the polynomial ring $\Z[T]$ via a morphism
defined in terms of the Chern--Schwartz--MacPherson (CSM) characteristic classes of
singular algebraic varieties. This morphism $\cF \to \Z[T]$ does {\em not} factor through
the usual Grothendieck ring of varieties $K_0(\cV_k)$, as we show explicitly in 
Example~\ref{ExB3Tri}. The main theorem showing the multiplicative property of 
this polynomial invariant
over disjoint unions of graphs is stated in Theorem \ref{mainhomo}, 
and its proof is reduced in
steps to a formula, given in Theorem~\ref{thmjoin}, for the 
Chern-Schwartz-MacPherson
classes of joins of disjoint subvarieties of projective space. 
In this same section, Proposition \ref{propsC} lists the main properties of the 
polynomial invariant, including the fact that it recovers as a special value of the derivative 
the usual Euler characteristic of the projective hypersurface complements, thus 
effectively correcting for its failure to be a Feynman rule. A way to compute the coefficients
of the polynomial invariant in terms of integrals of differential forms with logarithmic poles 
on a resolution is given in Remark~\ref{anInt}. The following section,  \S \ref{RenSec}, 
discusses the BPHZ renormalization procedure, in the formulation of the Connes--Kreimer
theory in terms of Birkhoff factorization of characters of the Hopf algebra of Feynman graphs.
Using the formulation in terms of characters with values in a Rota--Baxter algebra of weight
$-1$, one can show that the BPHZ procedure can be applied to the various cases of algebro-geometric 
Feynman rules considered here. In \S \ref{ZSec} we draw some analogies between the
partition functions obtained by summing over graphs the algebro-geometric Feynman rules
and the motivic zeta functions considered in the theory of motivic integration. Finally, \S 
\ref{joinpf} is devoted to the proof of Theorem  \ref{thmjoin}. 
The main ingredients
in the proof are Kwieci\'nski's product formula and Yokura's Riemann--Roch 
theorem for CSM classes, together with a blow-up construction.

\subsection{Feynman rules in quantum field theory}\label{Friqft}

The Feynman rules prescribe that, in perturbative scalar quantum field theory, one assigns to 
a Feynman graph a formal (usually divergent) integral 
\begin{equation}\label{UGamma}
U(\Gamma,p) = C\int \frac{\delta(\sum_{i=1}^n \epsilon_{v,i} k_i 
+ \sum_{j=1}^N \epsilon_{v,j} p_j)}{ q_1(k_1)\cdots q_n(k_n) } \, \frac{d^Dk_1}{(2\pi)^D} 
\cdots \frac{d^D k_n}{(2\pi)^D},
\end{equation}
where $C=\prod_{v\in V(\Gamma)} \lambda_v (2\pi)^D$, with $\lambda_v$ the coupling
constant of the monomial in the Lagrangian of degree equal to the valence of the vertex $v$.
Here, $n=\# E_{int}(\Gamma)$, and $N=\#E_{ext}(\Gamma)$.
The matrix $\epsilon_{v,i}$ is the incidence matrix of the (oriented) graph with entries 
$\epsilon_{v,i}=\pm 1$ if the edge $e_i$ is incident to the vertex $v$, outgoing or ingoing, and zero
otherwise. The $q_i(k_i)$ are the {\em Feynman propagators}. The latter are quadratic forms, given
in Euclidean signature by
\begin{equation}\label{propag}
q_i(k_i)= k_i^2 + m^2,
\end{equation}
where $k_i \in \R^D$ is the momentum variable associated to the edge $e_i$ of the graph,
with $k_i^2 =\| k_i \|^2$ the Euclidean square norm in $\R^D$ and $m\geq 0$ the mass
parameter. The integral $U(\Gamma,p)$ is a function of the external momenta $p=(p_e)_{e\in E_{ext}(\Gamma)}$, where the $p_e\in \R^D$ satisfy the conservation law $\sum_{e\in E_{ext}(\Gamma)} p_e =0$. The delta function in the numerator of \eqref{UGamma} imposes linear relations at each vertex
between the momentum variables, so that momentum conservation is preserved at each vertex.
This reduces the number of independent variables of integration from the number of edges to 
the number of loops.

The form of the Feynman integral \eqref{UGamma} immediately implies a multiplicative property.
Namely, if the Feynman graph is a disjoint union $\Gamma =\Gamma_1\cup \Gamma_2$ of two
components, then the integral satisfies
\begin{equation}\label{multiplUGamma}
U(\Gamma,p)=U(\Gamma_1,p_1)U(\Gamma_2,p_2),
\end{equation}
where $p_i$ are the external momenta of the graphs $\Gamma_i$, with $p=(p_1,p_2)$. This
follows from the fact that there are no linear relations between the momentum variables assigned
to edges in different connected components of the graph, so the integral splits as a product.

In quantum field theory one usually assembles the Feynman integrals of different graphs in
a formal series that gives, for fixed external momenta $p=(p_e)=(p_1,\ldots,p_N)$, the 
Green function
\begin{equation}\label{PertSer}
G(p)= \sum_\Gamma \frac{U(\Gamma, p)}{\# \Aut(\Gamma)},
\end{equation}
where $\Aut(\Gamma)$ are the symmetries of the graph. For a graph with 
several connected components, the symmetry factor behaves like 
\begin{equation}\label{symmprod}
\# \Aut(\Gamma) = \prod_j (n_j)! \prod_j \# \Aut(\Gamma_j)^{n_j},
\end{equation}
where the $n_j$ are the multiplicities (\ie there are $n_j$ connected components 
of $\Gamma$ all isomorphic to the same graph $\Gamma_j$). Thus, one can simplify
the combinatorics of graphs in quantum field theory by considering only connected
graphs and the corresponding connected Green functions.

One can further reduce the class of graphs that need to be considered, by passing to
the 1PI effective action, where only the graphs that are ``one-particle-irreducible" (1PI)
are considered. These are the two--edge--connected graphs, namely those that cannot
be disconnected by removal of a single edge. The reason why these suffice is again
related to the multiplicative properties of Feynman rules. A connected graph $\Gamma$ can be
described as a tree $T$ in which at the vertices one inserts 1PI graphs $\Gamma_v$ with number of
external edges equal to the valence of the vertex. The corresponding Feynman
integral can then be written in the form of a product 
\begin{equation}\label{UGamma1PI}
U(\Gamma,p) = \prod_{v\in T} U(\Gamma_v,p_v) \frac{1}{q_e((p_v)_e)} \delta((p_v)_e -(p_{v'})_e),
\end{equation}
\ie a product of Feynman integrals for 1PI graphs and inverses of the propagators $q_e$ 
for the edges of the tree, with momenta matching the external momenta of the 1PI graphs.

When one takes the dimensional regularization of the Feynman integrals, one replaces the
formal $U(\Gamma, p)$ by Laurent series, while maintaining the multiplicative properties
over disjoint unions of graphs. Thus, if one defines a polynomial ring $\cH$ generated by the 1PI
graphs with the product corresponding to the disjoint union, the dimensionally regularized
Feynman integral defines a ring homomorphism from $\cH$
to the ring $\cR$ of convergent Laurent series.
When the polynomial ring $\cH$ on the 1PI graphs is endowed with the Connes--Kreimer coproduct
as in \cite{CoKr}, and the ring of convergent Laurent series is endowed with the Rota--Baxter
operator $\fT$ of projection onto the polar part, one can implement the BPHZ renormalization of
the Feynman integral as in the Connes--Kreimer theory \cite{CoKr} as the Birkhoff factorization
of the Feynman integrals $U(\Gamma, p)$ into a product of ring homomorphisms from $\cH$
to $\fT\cR$ and $(1-\fT) \cR$, respectively defining the counterterms and the renormalized
part of the Feynman integral. 

In the following section we show how to abstract this setting to define 
algebro--geometric Feynman rules.

\section{Feynman rules in algebraic geometry}\label{FRAGSec}

We give an abstract definition of Feynman rules which encompasses the case of Feynman
integrals recalled above and that allows for algebro-geometric variants. 

\begin{defn}\label{FeyRuleDef}
A Feynman rule is an assignment of an element $U(\Gamma)$ in a commutative ring $\cR$
for each finite graph $\Gamma$, with the property that, for a disjoint union $\Gamma = \Gamma_1 \cup \cdots \cup \Gamma_k$ of connected graphs $\Gamma_i$, the function behaves multiplicatively
\begin{equation}\label{UGammaprodi}
U(\Gamma)=U(\Gamma_1) \cdots U(\Gamma_k).
\end{equation}
One also requires that, for a connected graph $\Gamma$ described as a finite
tree $T$ with vertices replaced by 1PI graphs $\Gamma_v$, the function $U(\Gamma)$ satisfies
\begin{equation}\label{1PIUGamma}
U(\Gamma) = U(L)^{\# E(T)} \prod_{v\in V(T)} U(\Gamma_v),
\end{equation} 
where $L$ is the graph consisting of a single edge.
Thus, a Feynman rule determines and is determined by a ring homomorphism
$U: \cH \to \cR$, where $\cH$ is the polynomial ring generated (over $\Z$) by the 1PI graphs
and by the assignment of the {\em inverse propagator} $U(L)$.
\end{defn}

The definition we give here, which will suffice for our purposes, covers the original Feynman 
integrals only in the case where one neglects the external momenta (or sets them all to zero) 
and remains with a nontrivial propagator for external edges given only by the mass $m^2$.
In fact, in that case, the formula \eqref{UGamma1PI} reduces to \eqref{1PIUGamma} with
$U(L)=1/m^2$ for all the external edges of the graphs $\Gamma_v$. The property 
\eqref{UGammaprodi} is the multiplicative property of the Feynman rules \eqref{multiplUGamma}.
The dimensionally regularized Feynman integral $U(\Gamma)$ is described then in terms 
of a ring homomorphism $U :\cH \to \cK$ to the ring of convergent Laurent series, by
identifying a monomial $\Gamma_1 \cdots \Gamma_k$ in $\cH$ with the disjoint union graph
$\Gamma = \Gamma_1 \cup\cdots \cup \Gamma_k$.

\smallskip

We are especially interested in algebro-geometric Feynman rules associated to
the parametric representation of Feynman integrals. In the parametric representation
for a massless theory, one reformulates the integral \eqref{UGamma} in the form
\begin{equation}\label{UGammaFeyPar}
U(\Gamma,p_1,\ldots,p_N) =C\, \frac{\Gamma(n-\frac{D\ell}{2})}{(4\pi)^{D\ell/2}} \int_{\sigma_n}
\frac{P_\Gamma(t,p)^{-n+D\ell/2}\, \omega_n}{\Psi_\Gamma(t)^{-n + (\ell+1)D/2}} ,
\end{equation}
where $t=(t_1,\ldots,t_n)\in \A^n$ with $n=\# E_{int}(\Gamma)$, integrated over the simplex
$\sigma_n =\{ t\in \R_+^n \,|\, \sum_i t_i =1 \}$, with volume form $\omega_n$ and with
$\Psi_\Gamma$ the Kirchhoff polynomial 
\begin{equation}\label{PsiGamma}
\Psi_\Gamma (t) = \det M_\Gamma(t), \ \ \ \text{ with } \ \ \ (M_\Gamma(t))_{rk} =\sum_i t_i \eta_{ir} \eta_{ik},
\end{equation}
where $\eta_{ik}$ is the circuit matrix of the graph (depending on a choice of orientation of the edges
$e_i$ and of a basis $\{ \ell_k \}$ of $H_1(\Gamma)$),
\begin{equation}\label{etamatr}
\eta_{ik} = \left\{\begin{array}{rl} +1 & \text{if edge } e_i \in \text{ loop } \ell_k, \text{ same orientation}\\
-1 &  \text{if edge } e_i \in \text{ loop } \ell_k, \text{ reverse orientation}\\
0 & \text{if edge } e_i \notin \text{ loop } \ell_k. \end{array} \right.
\end{equation}
(This is equivalent to the definition given in the introduction.)
If $b_1(\Gamma)=0$, we take $\Psi_{\Gamma}(t)=1$.

The function $P_\Gamma(t,p)$ is a homogeneus polynomial in $t$ of degree 
$b_1(\Gamma)+1$,
which also has a definition in terms of the combinatorics of the graph.
Notice that one can define
parametric representations in the case of massive theories $m\neq 0$ as well and obtain a
formulation similar to \eqref{UGammaFeyPar}, 
\begin{equation}\label{mUGammaFeyPar}
U(\Gamma,p_1,\ldots,p_N) =C\, \frac{\Gamma(n-\frac{D\ell}{2})}{(4\pi)^{D\ell/2}} \int_{\sigma_n}
\frac{V_\Gamma(t,p)^{-n+D\ell/2}\, \omega_n}{\Psi_\Gamma(t)^{D/2}} ,
\end{equation}
where, however, $V_\Gamma(t,p)$ is no longer a homogeneous polynomial in $t$.  Our definition
of Feynman rules in Definition \ref{FeyRuleDef} is modeled on the massive case, because of
the propagators $U(L)$ in \eqref{1PIUGamma}. 
In both the massive and the massless case, 
at least for sufficiently large spacetime dimension $D$, in the range where $n\leq D\ell/2$, the integral lives naturally on the complement in $\A^n$ of the affine hypersurface
\begin{equation}\label{hatXGamma}
\hat X_\Gamma = \{ t\in \A^n \,|\, \Psi_\Gamma(t) =0 \}.
\end{equation}

In the massless case where both $\Psi_\Gamma$ and $P_\Gamma(t,p)$ in \eqref{UGammaFeyPar} 
are homogeneous polynomials, one usually
reformulates the Feynman integral in terms of projective varieties and considers the complement
$\P^{n-1}\smallsetminus X_\Gamma$ of the projective hypersurface
\begin{equation}\label{XGamma}
X_\Gamma = \{ t=(t_1:\cdots: t_n) \in \P^{n-1} \,|\, \Psi_\Gamma(t)=0\}, 
\end{equation}
of which $\hat X_\Gamma$ is the affine cone. Although working in the projective setting is
very natural (see \cite{Blo}, \cite{BEK}), the discussion above indicates that, if one wants to
accommodate both massless and massive theories, it is more natural to work in the affine setting. Moreover, we will see here that working with the affine hypersurfaces
is better also from the point of view of having motivic Feynman rules.

\begin{defn}\label{AGFeynRules}
An algebro--geometric Feynman rule is an invariant $\bU(\Gamma)=\bU(\A^n\smallsetminus \hat X_\Gamma)$ of the graph hypersurface
complement, with values in a commutative ring $\cR$, with the following properties.
\begin{itemize}
\item For a disjoint union of graphs $\Gamma =\Gamma_1 \cup \Gamma_2$, it satisfies 
$\bU(\Gamma)=\bU(\Gamma_1)\bU(\Gamma_2)$.
\item For a connected graph $\Gamma$ obtained from a finite tree $T$ and 1PI graphs $\Gamma_v$
at the vertices $v\in V(T)$, it satisfies $\bU(\Gamma)=\bU(L)^{\# E(T)} \prod_{v\in V(T)} \bU(\Gamma_v)$,
where $\bU(L)$ is the value on the line $L$, \ie on the graph consisting of a single edge. 
\end{itemize}
An algebro--geometric Feynman rule is motivic if the invariant $\bU(\Gamma)$ 
only depends on the class $[\A^n \smallsetminus \hat X_\Gamma]$ of the hypersurface 
complement in the Grothendieck ring of varieties $K_0(\cV_\Q)$.
\end{defn}

By this definition, in particular, an algebro-geometric Feynman rule defines a
ring homomorphism $\bU:\cH \to \cR$ as in Definition \ref{FeyRuleDef}, by interpreting 
a monomial $\Gamma_1 \cdots \Gamma_k$ as the disjoint union 
$\Gamma = \Gamma_1 \cup\cdots \cup \Gamma_k$. In the motivic case this homomorphism 
factors through the commutative ring $K_0(\cV_\Q)$.

The dependence $\bU(\Gamma)=\bU(\A^n\smallsetminus \hat X_\Gamma)$ of an algebro-geometric
Feynman rule on the affine hypersurface complement should be understood here as a dependence
on the variety $\A^n \smallsetminus \hat X_\Gamma$ considered modulo linear changes of coordinates
in $\A^n$. This will be explained more in detail in \S \ref{UnivSec} below. It will be then be clear from
Lemma \ref{projaff} that, unlike the case of general Feynman rules, the second property
$\bU(\Gamma)=\bU(L)^{\# E(T)} \prod_{v\in V(T)} \bU(\Gamma_v)$ in Definition \ref{AGFeynRules}
will in fact be a consequence of the multiplicativity $\bU(\Gamma)=\bU(\Gamma_1)\bU(\Gamma_2)$
over disjoint unions $\Gamma =\Gamma_1 \cup \Gamma_2$, together with the fact that the
hypersurface complement does not distinguish between the case of the disjoint union $\Gamma =\Gamma_1 \cup \Gamma_2$ and the case where the graphs $\Gamma_1$ and $\Gamma_2$
are joined at a single vertex. 

Notice, moreover, that there are examples of combinatorially inequivalent connected 1PI graphs that
have the same graph hypersurface, so that one can construct Feynman rules that are 
not algebro-geometric or motivic, by assigning different invariants to such graphs, so that 
the resulting ring homomorphism $\cH \to \cR$ does not factor through $K_0(\cV_k)$ or through
the ring $\cF$ described in \S \ref{UnivSec} below.

\subsection{A universal algebro--geometric Feynman rule}\label{UnivSec}

We show that algebro--geometric Feynman rules, in the sense of Definition \ref{AGFeynRules},
correspond to ring homomorphisms from a universal ring $\cF$ to a given commutative ring. 
In particular, this defines a universal algebro--geometric Feynman rule obtained by assigning
$\bU(\Gamma)$ as the class of the hypersurface complement $\A^n \smallsetminus \hat X_\Gamma$
in the ring $\cF$. A motivic Feynman rule is then obtained by mapping $\cF$ to the Grothendieck ring
$K_0(\cV_\Q)$. 

\smallskip

We begin by the following simple observation, which explains why it is more convenient to work 
in the affine rather than the projective setting. 

\begin{lem}\label{projaff}
For every graph $\Gamma$, let $X_\Gamma\subset \P^{n-1}$ be the projective 
hypersurface \eqref{XGamma} and $\hat X_\Gamma \subset \A^n$ be its affine 
cone \eqref{hatXGamma}, with $n=\# E_{int}(\Gamma)$, as above. 

Let $\Gamma=\Gamma_1\cup \Gamma_2$ be the union of two disjoint graphs.
Then
\begin{equation}\label{AffProd}
\A^{n_1+n_2}\smallsetminus \hat X_\Gamma = (\A^{n_1} \smallsetminus \hat X_{\Gamma_1})\times
(\A^{n_2}\smallsetminus \hat X_{\Gamma_2}),
\end{equation}
where $n_i=\# E_{int}(\Gamma_i)$.

If neither $\Gamma_1$ nor $\Gamma_2$ is a forest, then the projective 
hypersurface complement $\P^{n_1+n_2-1} \smallsetminus 
X_\Gamma$ is a $\bG_m$-bundle over the product $(\P^{n_1-1}
\smallsetminus X_{\Gamma_1})\times (\P^{n_2-1}\smallsetminus 
X_{\Gamma_2})$ of the hypersurface complements of $\Gamma_1$ and 
$\Gamma_2$.

The same formulas hold if $\Gamma$ is obtained by attaching two disjoint
graphs $\Gamma_1$, $\Gamma_2$ at a vertex.
\end{lem}

\proof 
It is clear from both the combinatorial definition recalled in the introduction,
and from the definition \eqref{PsiGamma} in terms of Kirchoff matrices $M_\Gamma(t)$,
that if $\Gamma=\Gamma_1\cup \Gamma_2$ is a disjoint union
(or if $\Gamma$ is obtained by attaching $\Gamma_1$ and $\Gamma_2$
at a vertex), then 
\[
\Psi_\Gamma(t_1,\ldots,t_n) =\Psi_{\Gamma_1}(t_1,\ldots,t_{n_1}) 
\Psi_{\Gamma_2}(t_{n_1+1}, \ldots, t_{n_1+n_2}).
\]
This says that $\hat X_{\Gamma_1\cup \Gamma_2}$ is the hypersurface in 
$\A^{n_1+n_2}$
obtained as the union
\[
(\hat X_{\Gamma_1}\times \A^{n_2})\cup
(\A^{n_1} \times \hat X_{\Gamma_2}),
\]
and formula \eqref{AffProd} for the hypersurface complement 
$\A^{n_1+n_2}\smallsetminus \hat X_\Gamma$ follows immediately.

In projective terms, $X_\Gamma$ is given by the union
of the cones $C^{n_2}(X_{\Gamma_1})$, $C^{n_1}(X_{\Gamma_2})$
in $\P^{n_1+n_2-1}$ over $X_{\Gamma_1}$ and $X_{\Gamma_2}$, with vertices 
$\P^{n_2-1}$ and $\P^{n_1-1}$, respectively. Here one views the $\P^{n_i-1}$ 
containing $X_{\Gamma_i}$ as skew subspaces in $\P^{n_1+n_2-1}$. 
A point in the complement of $X_{\Gamma_1\cup \Gamma_2}$ in $\P^{n_1+n_2-1}$ 
is of the form
\[
(t_1: \cdots : t_{n_1+n_2}), \ \ \ \ 
\text{where} \ \ \  \Psi_{\Gamma_1}(t_1, \cdots , t_{n_1})\ne 0 \ \ \text{and} \ \  
\Psi_{\Gamma_2}(t_{n_1+1}, \cdots , t_{n_1+n_2})\ne 0.
\]
Note that  if $\Psi_{\Gamma_1}\not\equiv 1$ and 
$\Psi_{\Gamma_2}\not\equiv 1$,
then $\Psi_{\Gamma_1}(t_1: \cdots : t_{n_1})\ne 0$ only if
$(t_1: \cdots : t_{n_1})\ne 0$, and 
$\Psi_{\Gamma_2}(t_{n_1+1}: \cdots : t_{n_1+n_2})\ne 0$ only if
$(t_{n_1+1}: \cdots : t_{n_1+n_2})\ne 0$. This says that if neither 
$\Gamma_1$ nor $\Gamma_2$ is a forest, then we have a {\em regular\/}
map
\begin{equation}\label{regmap}
(\P^{n_1+n_2-1}\smallsetminus X_{\Gamma_1\cup \Gamma_2})
\to (\P^{n_1-1}\smallsetminus X_{\Gamma_1})
\times (\P^{n_2-1}\smallsetminus X_{\Gamma_2})
\end{equation}
given by 
\[
(t_1: \cdots : t_{n_1}:t_{n_1+1}: \cdots : t_{n_1+n_2}) \mapsto 
((t_1: \cdots : t_{n_1}),(t_{n_1+1}: \cdots : t_{n_1+n_2}))\quad.
\]
This map is evidently surjective, and the fiber over
$((t_1: \cdots : t_{n_1}),(t_{n_1+1} : \cdots : t_{n_1+n_2}))$
consists of the points
\[
(ut_1: \cdots : ut_{n_1}:vt_{n_1+1}: \cdots : vt_{n_1+n_2})
\]
with $(u: v)\in \P^1$, $uv\ne 0$. These fibers are tori $\bG_m(k)=k^*$,
completing the proof.

If either $\Gamma_1$ or $\Gamma_2$ is a forest, the corresponding
hypersurface $X_{\Gamma_i}$ is empty; the map \eqref{regmap} 
is not defined everywhere in this case.
\endproof

The observation above implies that, if we want to construct  Feynman rule 
$\bU(\Gamma)$
in terms of the hypersurface complements, then by working in the affine setting 
it suffices to have an invariant of affine varieties that is multiplicative on products 
and behaves in the natural way with respect to complements, that is, it 
satisfies an inclusion--exclusion property.
This indicates that the natural target of algebro-geometric Feynman rules
should be a ring reminiscent of the Grothendieck ring of varieties $K_0(\cV_k)$.
However, it will be advantageous to work in a ring with a more rigid 
equivalence relation than in the definition of $K_0(\cV_k)$: this will be
a ring mapping to $K_0(\cV_k)$, but also carrying enough information to
allow us to define Feynman rules by means of characteristic classes of
immersed varieties. The natural receptacle of our algebro-geometric Feynman 
rules will be the {\em Grothendieck ring of immersed conical varieties}, which 
we define as follows.

\begin{defn}\label{GrothImm}
Let $\cF$ be the ring whose elements are formal finite integer linear combinations of 
equivalence classes of closed
conical (that is, defined by homogeneous ideals) reduced algebraic sets $V$
of $\A^\infty$, such that $V\subseteq \A^N$ for some finite $N$,
modulo the equivalence relation given by linear changes of coordinates,
and with the further relation dictating `inclusion-exclusion':
\begin{equation}\label{inclexclF}
[V\cup W] = [V]+[W] - [V \cap W] \quad.
\end{equation}
The ring structure is given by the product induced by
\begin{equation}\label{prodform}
[V]\cdot [W] = [V \times W]\quad.
\end{equation}
\end{defn}

This is an embedded version of the Grothendieck ring,
and it maps to the Grothendieck ring since varieties differing by a linear change 
of coordinates 
are isomorphic. It will also map to polynomial rings, via characteristic classes
of complements, as we will explain in~\S \ref{CharSec}. 

If $U$ is a locally closed set, defined as the complement $V\smallsetminus
W$ of two closed conical subsets, we can define a class
\[
[U]:=[V]-[W]
\]
in $\cF$; the inclusion--exclusion property guarantees that this assignment
is independent of the specific representation of $U$, and that the product
formula \eqref{prodform} extends to locally closed sets.
This implies that ring homomorphisms $\cF \to \cR$ of the Grothendieck
ring of immersed conical varieties to arbitrary commutative rings define algebro-geometric Feynman rules:

\begin{prop}\label{masterlem}
Let $I: \cF \to \cR$ be a ring homomorphism to a commutative ring $\cR$. 
For every graph $\Gamma$ with $n=\# E_{int}(\Gamma)$, define 
$\bU(\Gamma)\in R$ by
\begin{equation}\label{IAIXGamma}
\bU(\Gamma):=I([\A^n])-I([\hat X_\Gamma])  = I([\A^n \smallsetminus \hat X_\Gamma]) \quad.
\end{equation}
Then $\bU$ is multiplicative under disjoint unions: if $\Gamma_1$, 
$\Gamma_2$ are disjoint graphs, then $\bU(\Gamma_1 \cup \Gamma_2)=
\bU(\Gamma_1)\cdot \bU(\Gamma_2)$.

The same formula holds if $\Gamma_1$, $\Gamma_2$ share a 
single vertex. Moreover, if $\Gamma$ is obtained by connecting two 
disjoint graphs $\Gamma_1$, $\Gamma_2$ by an edge, then the invariant 
satisfies 
\begin{equation}\label{prodedge}
\bU(\Gamma) = \bU(\Gamma_1) \bU(L) \bU(\Gamma_2),
\end{equation}
where $\bU(L)$ is the invariant associated to the graph $L$ consisting of a single edge.
This is given by $\bU(L)=I([\A^1])=:\fL$, the value of $I$ on the class of the affine line.
\end{prop}

\proof
The claims are all preserved under homomorphisms, so it suffices to
prove them for the invariant $\bU$ with values in $\cF$ defined by
\[
\bU(\Gamma):= [\A^n \smallsetminus \hat X_\Gamma]\in \cF
\]
for a graph $\Gamma$ with $n$ internal edges. The multiplicativity
under disjoint unions, and under the operation of attaching graphs
at a single vertex, follows then immediately from formula \eqref{AffProd}
in Lemma~\ref{projaff}. In turn, formula~\eqref{prodedge} follows from
the multiplicativity. To see that $\bU(L)=[\A^1]$, simply recall that
the graph hypersurface corresponding to a single edge (or to any
forest) is~$\emptyset$.
\endproof

We denote here by $\fL$ the value $I([\A^1])$, as this will map to the 
Lefschetz motive $\bL$ in the Grothendieck group. Note that we then 
have $I([\A^n])=\fL^n$; by definition, this is the invariant associated
with any forest with $n$ edges, since the graph polynomial of a 
forest is~$1$ (and hence the corresponding graph hypersurface
is empty). In particular, $1=I([\A^0])=\bU(\star)$ is the invariant of 
the trivial graph $\star$ consisting of one vertex and no edges.
\smallskip

We have given in Definition \ref{FeyRuleDef} an equivalent
characterization of Feynman rules in terms of a ring homomorphism
$\bU: \cH \to \cR$ together with an ``inverse propagator" $\bU(L)$. 
An algebro-geometric Feynman rule defined as above by a ring 
homomorphism $I: \cF \to \cR$ corresponds, in these terms, to the 
homomorphism $\bU: \cH \to \cR$ obtained by precomposition with the 
ring homomorphism
$$ \cH \to \cF,  \ \ \  \Gamma\mapsto [\A^n] - [\hat X_\Gamma], $$
for all connected 1PI graphs $\Gamma$, and with the inverse propagator 
$[\A^1]\in \cF$.
\[
\xymatrix@R=7pt@C=7pt{
& & [\A^n]-[\hat X_\Gamma] \ar@/^1pc/@{|->}[dddddrr]\\
& &  \cF  \ar[dddr]^I \\
\\
\\
& \cH \ar[uuur] \ar[rr]_\bU & & \cR \\
\qquad\Gamma\qquad \ar@/_1pc/@{|->}[rrrr] 
\ar@/^1pc/@{|->}[uuuuurr] & & & & I([\A^n]) - I([\hat X_\Gamma])
}
\]
\bigskip

By Proposition \ref{masterlem}, we have a `universal' algebro-geometric 
Feynman rule given by the identity homomorphism $I: \cF \to \cF$. Again,
this corresponds to the ring homomorphism $\cH \to \cF$ that assigns 
$[\A^n] - [\hat X_\Gamma]$ to a connected 1PI graph with 
$n=\# E_{int}(\Gamma)$ and with inverse propagator $[\A^1]\in \cF$.

\subsection{Operations on graphs and Feynman rules}\label{OpsSec}

The universal algebro--geometric Feynman rule defined by $[\A^n] - 
[\hat X_\Gamma]$ in $\cF$
satisfies the following properties for elementary geometric operations on a 
graph. These properties are inherited by any other algebro-geometric 
Feynman rule.

\begin{itemize}
\item Let $\Gamma'$ be obtained from $\Gamma$ by attaching an edge to
a vertex of $\Gamma$. Then
\[
\bU(\Gamma')=[\A^1]\cdot \bU(\Gamma)\quad.
\]
\item Let $\Gamma$ be a graph that is not 1PI. Then $\bU(\Gamma)$ is of the form
$$ \bU(\Gamma)=[\A^1] \cdot B $$ for some class $B\in \cF$.
\item Let $\Gamma'$ be obtained from $\Gamma$ by splitting an edge.
Then
\[
\bU(\Gamma')=[\A^1] \cdot \bU(\Gamma)\quad.
\]
\item Let $\Gamma'$ be obtained from $\Gamma$ by attaching a looping 
edge to a vertex. Then
\[
\bU(\Gamma')=([\A^1]-1) \cdot \bU(\Gamma)\quad.
\]
\item Let $\Gamma$ be an $n$-side polygon. Then
\[
\bU(\Gamma)=[\A^n]- [\A^{n-1}]\quad.
\]
\end{itemize}

All these properties follow very easily from the definition of $\bU(\Gamma)$. For instance,
the property for non-1PI graphs follows directly from \eqref{prodedge}, while 
attaching a looping edge amounts to
multiplying the equation of the graph hypersurface by a new variable,
and viewing the result in a space of dimension~$1$ higher.
In affine space, and in terms of the complement, this is clearly the
same as taking a product by $\A^1\smallsetminus \A^0$.

\subsection{Motivic Feynman rules}\label{MotFrules}

The ring $\cF$ maps to the Grothendieck ring of varieties $K_0(\cV_k)$ by mapping the
equivalence class $[X]\in \cF$ under linear coordinate changes to the isomorphism class $[X] \in K_0(\cV_k)$. It is a ring homomorphism since the product is in both cases defined by the class of the
product of manifolds. Thus, one obtains in this way a motivic Feynman rule defined by 
$\Gamma\mapsto [\A^n \smallsetminus \hat X_\Gamma]\in K_0(\cV_k)$. 
This corresponds to
the ring homomorphism $\bU:\cH \to K_0(\cV_k)$ that maps the monomial
$\Gamma_1 \cdots \Gamma_k$, where the $\Gamma_i$ are 1PI graphs, to the class
\begin{equation}\label{motFR}
\bU(\Gamma_1 \cdots \Gamma_k)= [\A^n \smallsetminus \hat X_\Gamma] =[\A^{n_1}\smallsetminus \hat X_{\Gamma_1}]\cdots [\A^{n_k}\smallsetminus \hat X_{\Gamma_k}], 
\end{equation}
where $\Gamma =\Gamma_1 \cup \cdots \cup \Gamma_k$ is the disjoint union and $n=\sum_i n_i$.
The inverse propagator is $\bU(L)=\bL=[\A^1]$, the Lefschetz motive, \ie the class of the affine line
in $K_0(\cV_k)$. 

This means that one can think of the ``propagator" as being the formal inverse $\bL^{-1}$ of the Lefschetz motive. This corresponds to the Tate motive $\Q(1)$ when one maps in the
natural way (see \cite{GS}) the Grothendieck ring of varieties $K_0(\cV_k)$ to the Grothendieck ring
of motives $K_0(\cM_k)$.

The relation between the motivic Feynman rule \eqref{motFR} and the hypersurface complement
in projective space is described as follows.

\begin{lem}\label{affprojL}
If $\Gamma$ is not a forest,
the hypersurface complements $\A^n\smallsetminus \hat X_{\Gamma}$ and 
$\P^{n-1}\smallsetminus X_\Gamma$ are related in the Grothendieck ring $K_0(\cV_k)$ by
\begin{equation}\label{affprojK0}
[ \A^n\smallsetminus \hat X_{\Gamma} ] = (\bL -1) \, [\P^{n-1} \smallsetminus X_\Gamma].
\end{equation}
\end{lem}

\proof We have 
\begin{equation}\label{hatXX}
[\hat X_\Gamma] = (\bL -1) [X_\Gamma] +1, 
\end{equation}
since $\hat X_\Gamma$ is the affine cone in $\A^n$ over $X_\Gamma$. Thus, we obtain
$$ \begin{array}{rl}
[ \A^n\smallsetminus \hat X_{\Gamma} ] = & \bL^n - 1 - (\bL-1)[X_\Gamma] \\[2mm] = &
(\bL-1) (\bL^{n-1}+\cdots+\bL+1 - [X_\Gamma]) \\[2mm] = & (\bL-1) ([\P^{n-1}]-[X_\Gamma]). 
\end{array} $$
\endproof

Thus, we see that the factor $(\bL-1)$ restores the multiplicative property of Feynman rules
that is not satisfied at the level of the projective hypersurface complements. 

\begin{ex} {\rm The graph hypersurfaces corresponding to the so-called
{\em banana graphs\/} are studied in \cite{AlMa}. Lemma~\ref{affprojL}
and formula (3.13) in \cite{AlMa} yield that
\[
[\A^n\smallsetminus \hat X_{\Gamma_n}]
=(\bL-1) \frac{(\bL-1)^n-(-1)^n}{\bL} + n (\bL-1)^{n-1}\quad,
\]
where $\Gamma_n$ denotes the $n$-th banana graph ($n$ parallel
edges joining two vertices).}
\end{ex}

Given the motivic Feynman rule determined by the ring homomorphism 
$\bU: \cH \to K_0(\cV_k)$,
with $\bU(\Gamma)=[\A^n\smallsetminus \hat X_\Gamma]$ and inverse
propagator $\bL$, one can obtain other motivic
Feynman rules with values in commutative rings $\cR$ using motivic measures. 
Recall that
a {\em motivic measure} is by definition a ring homomorphism $\mu : K_0(\cV_k) \to \cR$ (see for
instance \cite{LaLu}, \S 1.3), so that the composite $\mu \circ \bU$ defines an $\cR$-valued
motivic Feynman rule.  

\smallskip

Notice that, in particular, when one considers the ring homomorphism from 
$K_0(\cV_k)$ to $\Z$
given by the ordinary topological Euler characteristic, the image of the classes 
$[\A^n\smallsetminus \hat X_\Gamma]$ 
is zero if $\Gamma$ is not a forest,
as one can see from the presence
of the torus factor $[\bG_m]=\bL-1$ in \eqref{affprojK0}, while if one computes the
Euler characteristic of the projective hypersurface complements 
$[\P^{n-1}\smallsetminus X_\Gamma]$ this will in general be non-zero (see for instance
the examples computed in \cite{AlMa}) but the multiplicative property of Feynman rules
is no longer satisfied. We show in \S \ref{CharSec} below how one can define an
algebro-geometric Feynman rule that assigns a polynomial invariant 
in $\Z[T]$ to the class in $\cF$ of each hypersurface complement 
$\A^n\smallsetminus \hat X_\Gamma$,
in such a way that the value at zero of the derivative of the polynomial recovers the
Euler characteristic of the complement of the
projective hypersurface $X_\Gamma$. This invariant will
be our best answer to the question of a generalization of the ordinary Euler
characteristic that satisfies the multiplicative property of Feynman rules and from which
the usual Euler characteristic can be recovered as a special value.
This invariant is not obtained from a homomorphism 
$K_0(\cV_k)\to \Z[T]$ as the following example shows. 

\begin{ex}\label{ExB3Tri}{\rm  The two graphs
\begin{center}
\includegraphics[scale = .4]{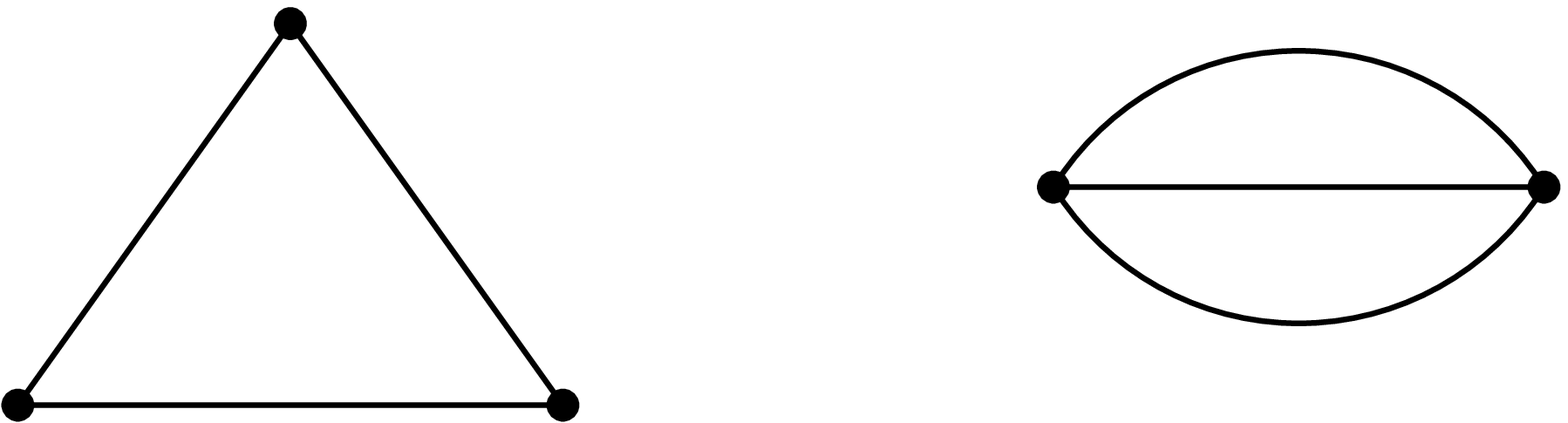}
\end{center}
have the same motivic invariant $[\A^3]-[\A^2]$, but different polynomial
invariants: $T(T+1)^2$ and $T(T^2+T+1)$, respectively. }
\end{ex}

\smallskip

It is proved in \cite{LaLu} that the quotient of the Grothendieck ring $K_0(\cV_\C)$ by the
ideal generated by $\bL=[\A^1]$ is isomorphic as a ring to $\Z[SB]$, the ring of the multiplicative
monoid $SB$ of stable birational equivalence classes of varieties in $\cV_\C$. Recall that two
(irreducible) varieties $X$ and $Y$ are stably birationally equivalent if $X\times \P^n$
and $Y\times \P^m$ are birationally equivalent for some $n, m\geq 0$. The observations
of \S  \ref{OpsSec} above then give the following.

\begin{prop}\label{SB1PI}
Let $\Gamma$ be a graph that is {\em not} 1PI. Then the stable birational equivalence class of the projective graph hypersurface satisfies $[X_\Gamma]_{sb}=1$ in $\Z[SB]$.
\end{prop}

\proof We know by Lemma \ref{affprojL} that, in the Grothendieck ring $K_0(\cV_\C)$, we have
$[\A^n\smallsetminus \hat X_\Gamma]= \bL^n-1 - (\bL-1)[X_\Gamma]$. Moreover, by the
observation made in \S \ref{OpsSec} we know that for a graph $\Gamma$ that is not 1PI the
class $[\A^n\smallsetminus \hat X_\Gamma]= \bL \cdot [\A^{n-1} \smallsetminus \hat X_{\Gamma'}]$,
where $\Gamma'$ is the graph obtained from $\Gamma$ by removing a disconnecting edge $L$
and $\bL=[\A^1]=\bU(L)$. Then we use the fact that $\Z[SB]=K_0(\cV_\C)/(\bL)$ as in \cite{LaLu},
and we obtain that $[\A^n\smallsetminus \hat X_\Gamma]_{sb}=0 \in \Z[SB]$, while $\bL^n-1 - (\bL-1)[X_\Gamma] \in K_0(\cV_\C)$ becomes $-1 + [X_\Gamma]_{sb} \in \Z[SB]$, so that we obtain $[X_\Gamma]_{sb}-1 =0 \in \Z[SB]$.
\endproof

\smallskip

A variant of the motivic Feynman rule \eqref{motFR} is obtained by setting
\begin{equation}\label{motFR2}
\bU(\Gamma)=\frac{ [\A^n \smallsetminus \hat X_\Gamma] }{\bL^n},
\end{equation}
with values in the ring $K_0(\cV_k)[\bL^{-1}]$, where one inverts the Lefschetz motive.
Dividing by $\bL^n$ has the effect of normalizing the ``Feynman integral" $\bU(\Gamma)$
by the value it would have if $\Gamma$ were a forest on the same number of edges. For the
original Feynman integrals this would measure the amount of linear dependence
between the edge momentum variables created by the presence of the interaction vertices.
We will discuss in \S \ref{RenSec} some advantages of using the motivic Feynman rule 
\eqref{motFR2} as opposed to \eqref{motFR}.

\smallskip

Moreover, notice that, modulo the important problem of divergences of the Feynman integral,
which needs to be treated via a suitable regularization and renormalization procedure, which 
in the algebro-geometric setting often involves blowups of the divergence locus (see \cite{BEK}),
one would like to think of the original Feynman rule given by the parametric Feynman integral
as an algebro-geometric Feynman rule with values in the algebra $\cP$ of periods.
Recall that conjecturally (see \cite{KoZa}) the algebra $\cP$ of periods is generated over $\Q$
by equivalence classes of the form $[(X,D,\omega,\sigma)]$, where $X$ is a 
smooth affine variety over $\Q$, $D\subset X$ is a normal crossings divisor, $\omega \in \Omega^{\dim(X)}(X)$ is an algebraic differential form, and $\sigma \in H_{\dim(X)} (X(\C),D(\C);\Q)$ is a relative
homology class. The equivalence relation is taken modulo the change of variables formula and the Stokes formula for integrals (see \cite{KoZa} for more details). In the setting that we are considering,
where in Feynman integrals we set the external momenta equal to zero and keep a non-zero mass,
so that the Feynman rules satisfy \eqref{UGammaprodi} and \eqref{1PIUGamma}, the function
$V_\Gamma(t,p)$ in the numerator of the parametric Feynman integral \eqref{mUGammaFeyPar}
is reduced to $V(t,p)|_{p=0}= m^2$. This follows from the fact that, in general, $V_\Gamma(t,p)$ 
is of the form
$$ V_\Gamma(t,p) = p^\dag R_\Gamma(t) p + m^2, $$
where $R_\Gamma(t)$ is another matrix associated to the graph $\Gamma$ defined in terms
of cut-sets, whose explicit expression we do not need here (the interested reader can see
for instance \cite{ItZu} or \cite{BjDr}). Thus, for the massive case with zero external
momenta, the parametric Feynman integral is, up to a multiplicative constant and a possibly
divergent $\Gamma$-factor, of the form
\begin{equation}\label{FeyParp0}
\int_{\sigma_n} \frac{\omega_n}{\Psi_\Gamma(t)^{D/2}} .
\end{equation}
Modulo the important issue of divergences coming from the nontrivial intersections 
$\sigma_n \cap \hat X_\Gamma$, we can then think of the original Feynman rule
as a morphism to the algebra of periods $\cP$ that assigns 
\begin{equation}\label{UGammaPer}
 \bU(\Gamma) =[( \A^n\smallsetminus \hat X_\Gamma, \hat\Sigma_n, \Psi_\Gamma^{-D/2} \omega_n, \sigma_n )], 
\end{equation} 
where $\hat\Sigma_n=\{ t\in \A^n\,|\, \prod_i t_i =0 \}$. 

A possible way to handle the divergences in terms of ``integrating around the singularities" using
Leray coboundaries was proposed in \cite{Mar}. We discuss briefly in \S \ref{RenSec} how this
might fit with Feynman rules of the form \eqref{UGammaPer}.

\section{Characteristic classes and Feynman rules}\label{CharSec}
In this section we define a ring homomorphism 
\[
I_{CSM}: \cF \to \Z[T]\quad,
\]
and hence (by Proposition~\ref{masterlem}) obtain a polynomial valued 
Feynman rule. We will denote by $C_\Gamma(T)$ the invariant
corresponding to $I_{CSM}$ for a graph $\Gamma$: that is,
\[
C_\Gamma(T):= I_{CSM}(\A^n)-I_{CSM}(\hat X_\Gamma)\quad,
\]
if $\Gamma$ has $n$ (internal) edges.

This invariant will carry information related to the 
Chern-Schwartz-MacPherson (CSM) class of the graph hypersurface
of a given graph~$\Gamma$. The reader is addressed to \S2.2 of
\cite{AlMa} for a quick review of the definition and basic properties 
of these classes. 

Before defining $I_{CSM}$, we highlight a few features of the 
invariant. 

\begin{prop}\label{propsC}
Let $\Gamma$ be a graph with $n$ edges.
\begin{itemize}
\item $C_\Gamma(T)$ is a monic polynomial of degree~$n$.
\item If $\Gamma$ is a forest, then 
$C_\Gamma(T)=(T+1)^n$. In particular, the inverse propagator
corresponds to $T+1$.
\item If $\Gamma$ is not a forest, then $C_\Gamma(T)$ is a multiple
of $T$.
\item The coefficient of $T^{n-1}$ in $C_\Gamma(T)$ equals 
$n-b_1(\Gamma)$.
\item The value $C_\Gamma'(0)$ of the derivative of $C_\Gamma(T)$
at $0$ equals the Euler characteristic of the complement $\P^n
\smallsetminus X_\Gamma$.
\end{itemize}
\end{prop}

The proof of this lemma will follow the statement of Theorem~\ref{mainhomo}.

Of course, the invariant will also satisfy the properties listed in \S\ref{OpsSec}.
These take the following form:
\begin{itemize}
\item Let $\Gamma'$ be obtained from $\Gamma$ by attaching an edge to
a vertex of $\Gamma$, or by splitting an edge of $\Gamma$. Then
\[
C_{\Gamma'}(T)=(T+1)\cdot C_\Gamma(T)\quad.
\]
\item Let $\Gamma'$ be obtained from $\Gamma$ by attaching a looping 
edge to a vertex. Then
\[
C_{\Gamma'}(T)=T \cdot C_\Gamma(T)\quad.
\]
\item Let $\Gamma$ be a graph that is not 1PI. Then 
$C_\Gamma(-1)=0$.
\item Let $\Gamma$ be an $n$-side polygon. Then
\[
C_\Gamma(T)=T(T+1)^{n-1}\quad.
\]
\end{itemize}

\begin{rem}{\rm
The parallel between the motivic invariant introduced in \S\ref{MotFrules}
is even more apparent if one changes the variable $T$ to $L=T+1$.
We choose $T$ because it has a compelling geometric interpretation:
$T^k$ corresponds to the class $[\P^k]$ in the ambient space which
we use to define the invariant. Ultimately, this is due to the fact that
the CSM class of a torus $\bT^k$ embedded in $\P^k$ as the complement
of the `algebraic symplex' is~$[\P^k]$, cf.~Theorem~4.2 
in~\cite{Alu1}.}
\end{rem}
\smallskip

In order to define $I_{CSM}$, it suffices to define it on a set of generators
of $\cF$, and verify that the definition preserves the relations defining $\cF$.

Generators for $\cF$ consist of conical subvarieties $\hat X\subseteq \A^N$.
View $\hat X$ as a locally closed subset of $\P^N$; as such, it determines 
a CSM class in the Chow group $A(\P^N)$ of $\P^N$:
\[
\csm(\one_{\hat X})=a_0 [\P^0]+\cdots + a_N[\P^N]\quad.
\]
(Here $\one$ denotes the constant function~$1$ on the specified locus;
we denote by $\csm$ MacPherson's natural transformation relating
constructible functions and classes in the Chow group.)

\begin{defn}\label{Gdef}
We define
\[
G_{\hat X}(T):=a_0 + a_1 T +\cdots + a_N T^N\quad.
\]
\end{defn}

\begin{ex}\label{GAN}{\rm
For $\hat X=\A^N$:
\[
G_{\A^N}(T):=(T+1)^N\quad.
\]
Indeed, viewing $\A^N$ as the complement of $\P^{N-1}$ in $\P^N$ gives
\[
\csm(\one_{\A^N})=\csm(\one_{\P^N})-\csm(\one_{\P^{N-1}})
=((1+H)^{N+1}-H (1+H)^N)\cap [\P^N]=(1+H)^N \cap [\P^N]\quad,
\]
where $H$ denotes the hyperplane class in $\P^N$. The coefficient
of $[\P^k]$ in this class is $\binom Nk$, with the stated result.
}\end{ex}

\begin{rem}{\rm
Here are a few comments on the definition of $G_{\hat X}(T)$.
\begin{itemize}
\item The definition does not depend on the dimension of the ambient
affine space $\A^N$: the largest $i$ for which $a_i\ne 0$ is the 
dimension of $\hat X$.
\item If $\hat X$ and $\hat X'$ differ by a coordinate change, then clearly
$G_{\hat X}(T)=G_{\hat X'}(T)$.
\item If $\hat X, \hat Y\subseteq \A^N$, then
\[
G_{\hat X\cup \hat Y}(T) = G_{\hat X}(T) + G_{\hat Y}(T) - G_{\hat X\cap \hat Y}(T)\quad:
\]
this follows from the inclusion-exclusion property of CSM classes.
\item By the previous two points, the definition goes through the equivalence
relation defining $\cF$. Thus, we can define a map $I_{CSM}: \cF \to
\Z[T]$ by setting
\[
I_{CSM}([\hat X]):= G_{\hat X}(T)\quad,
\]
and extending by linearity. This map is clearly a group homomorphism.
\end{itemize}}
\end{rem}
\smallskip

We claim that:

\begin{thm}\label{mainhomo}
$I_{CSM}$ is a homomorphism {\em of rings.\/}
\end{thm}

Once Theorem~\ref{mainhomo} is established, Proposition~\ref{masterlem}
will show that setting
\[
C_\Gamma(T)=U_{CSM}(\Gamma):=I_{CSM}([\A^n])-I_{CSM}([\hat X_\Gamma])
=G_{\A^n}(T)-G_{\hat X_\Gamma}(T)\quad,
\]
where $n=$ the number of edges of $\Gamma$, defines a multiplicative graph
invariant. The polynomial $C_\Gamma(t)$ satisfies all the properties
listed at the beginning of this subsection:

\proof[Proof of Proposition~\ref{propsC}]
Since $\hat X_\Gamma$ is properly contained in $\A^n$, the dominant term in the
difference $G_{\A^n}(T)-G_{\hat X_\Gamma}(T)$ is $T^n$: this proves the
first point.

If $\Gamma$ is a forest, then $\hat X_{\Gamma}=\emptyset$. Thus
$C_\Gamma(T)=G_{\A^n}(T)=(T+1)^n$ (Example~\ref{GAN}), proving
the second point.

If $\Gamma$ is not a forest, $\hat X_{\Gamma}\ne \emptyset$, and the
Euler characteristic of $\hat X_{\Gamma}$ is $1$ ($\hat X_{\Gamma}$ is
an affine cone). Therefore the constant term of $G _{\A^n}(T)-
G_{\hat X_\Gamma}(T)$ is $1-1=0$: this proves that $C_{\Gamma}(T)$
is a multiple of $T$ in this case, as claimed.

As to the fourth point: if $\Gamma$ is a forest, then $b_1(\Gamma)=0$
and the formula is verified. Thus, assume $\Gamma$ is not a forest.
The coefficient of $T^{n-1}$ in $G _{\A^n}(T)=(T+1)^n$ is $n$, while
the coefficient of $T^{n-1}$ in $G _{\hat X_\Gamma}(T)$ equals
the coefficient of the top-dimensional term $[\P^{n-1}]$ in 
$\csm(\hat X_\Gamma)$. This equals the degree of the hypersurface
$X_\Gamma$, which is $b_1(\Gamma)$, and the formula follows.

Finally, $C_\Gamma'(0)$ equals the coefficient of $T$ in $C_\Gamma(T)$.
If $\Gamma$ is a forest, then $C_\Gamma(T)=(T+1)^n$, so this 
coefficient is $n$, and equals the Euler characteristic of $\P^{n-1}\smallsetminus
\emptyset$. If $\Gamma$ is not a forest, then the coefficient of $T$ in
$C_\Gamma(T)$ equals the coefficient of $[\P^0]$ in the 
Chern-Schwartz-MacPherson class of $\P^{n-1} \smallsetminus X_\Gamma$
(see Proposition~\ref{notfore}, below). This equals the Euler characteristic
of $\P^{n-1} \smallsetminus X_\Gamma$, by general properties of
Chern-Schwartz-MacPherson classes (see for example \cite{AlMa}, 
\S2.2).
\endproof

As in the motivic case, the invariant $C_\Gamma(T)$ can be expressed 
in terms of the complement of the projective graph hypersurface. The analog
of Lemma~\ref{affprojL} is:

\begin{prop}\label{notfore}
If $\Gamma$ is not a forest, and has $n$ edges, then
\[
\csm(\one_{\P^{n-1}\smallsetminus X_\Gamma})=H^n C_\Gamma(1/H)
\cap [\P^{n-1}]\quad,
\]
where $H$ is the hyperplane class in $\P^{n-1}$.
Otherwise put, if $\Gamma$ is not a forest, then $C_\Gamma(T)$ may 
be recovered from $\csm(\one_{\P^{n-1}\smallsetminus X_\Gamma})$
by replacing the class
$[\P^k]$ in $\csm(\one_{\P^{n-1}\smallsetminus X_\Gamma})$ by 
$T^{k+1}$.
\end{prop}

\begin{proof}
Indeed, if $\csm(X)=f(H)\cap [\P^{n-1}]$, then
\begin{equation}\label{csmaff}
\csm(\one_{\A^n\smallsetminus {\hat X}_\Gamma})
=\csm(\A^n)- \csm({\hat X}_\Gamma)
=((1+H)^n - f(H) - H^n)\cap [\P^n]\quad:
\end{equation}
this follows from a straightforward computation, using the formula for
the CSM class of a cone (Proposition~5.2 in \cite{AlMa}).
Formula \eqref{csmaff} says that $H^n C_\Gamma(1/H)=(1+H)^n-f(H)-H^n$.
On the other hand, the polynomial $(1+H)^n - H^n - f(H)$ corresponds
to $\csm(\one_{\P^{n-1}\smallsetminus X_\Gamma})$
in $\P^{n-1}$; this is precisely the statement.
\end{proof}

\begin{ex} {\rm For $\Gamma_n=$ the $n$-edge banana graph,
\[
C_{\Gamma}(T) =T(T-1)^{n-1} + n T^{n-1}\quad.
\]
Indeed, Remark~4.11 in \cite{AlMa} gives
\[
\csm(\one_{\P^{n-1}\smallsetminus X_\Gamma})=((1-H)^{n-1}+nH)
\cap [\P^{n-1}]\quad.
\]
By Proposition~\ref{notfore}, therefore,
\[
H^n C_\Gamma(1/H) = (1-H)^{n-1}+nH
\]
and the result follows at once.}
\end{ex}

\begin{rem}\label{anInt}{\rm
Let $\Gamma$ be a graph with $n$ edges, that is not a forest, and
suppose 
\[
C_\Gamma(T)=T^n + a_{n-2} T^{n-1} + \cdots + a_0 T\quad.
\]
Let $\pi: \widetilde \P^{n-1} \to \P^{n-1}$ be a proper birational map
such that $D:=\pi^{-1}(X_{\Gamma})$ is a divisor with normal crossings
and nonsingular components. Then
\[
a_k=\int (\pi^* H)^k \cdot c(T_{\widetilde\P^{n-1}}(-\log D))\cap 
[\widetilde \P^{n-1}]\quad,
\]
where $T_{\widetilde\P^{n-1}}(-\log D)$ denotes the dual of the bundle 
$\Omega^1_{\widetilde\P^{n-1}}(\log D)$
of differential forms with logarithmic poles along $D$, and $\int \alpha$ stands
for the degree of the class $\alpha$, in the sense of \cite{Ful}, Definition~1.4.

This follows immediately from Proposition~\ref{notfore} and the 
expression of $\csm(\one_{\P^{n-1}\smallsetminus \hat X_\Gamma})$
in terms of Chern classes of logarithmic forms (cf.~\cite{AlMa}, \S2.3).
}\end{rem}

\smallskip

In the rest of this section we reduce the proof of Theorem~\ref{mainhomo}
to a statement (Theorem~\ref{thmjoin}) concerning Chern-Schwartz-MacPherson 
classes of joins of disjoint subvarieties of projective space. The proof of this
statement will be deferred to \S\ref{joinpf}.

In order to prove Theorem~\ref{mainhomo}, it suffices to prove that
\[
G_{\hat X\times \hat Y}(T)=G_{\hat X}(T)\cdot G_{\hat Y}(T)
\]
for all conical affine varieties $\hat X\subseteq\A^m$, $\hat Y\subseteq \A^n$.
If $\hat X=\emptyset$ or $\hat Y=\emptyset$, this is immediate, as this identity
is $0=0$ in this case.
If $\hat X$ or $\hat Y$ is a point (that is, the origin of the ambient affine space), 
the identity is also immediate. Indeed, $G_{\A^0}(T)=1$.

Therefore:

\begin{lem}\label{lemmahomo}
In order to prove Theorem~\ref{mainhomo}, it suffices to prove that
if $\hat X\subseteq \A^m$, resp.~$\hat Y\subseteq \A^n$ are affine cones 
over projective algebraic sets $X\subseteq \P^{m-1}$, 
resp.~$Y\subseteq \P^{n-1}$, then
\[
G_{\hat X\times \hat Y}(T)=G_{\hat X}(T)\cdot G_{\hat Y}(T)\quad.
\]
\end{lem}

What is a little surprising now is that this is {\em not\/} obvious. There is
a `product formula for CSM classes', due to Kwieci\'nski 
(\cite{Kwi}, \cite{Alu1}); but it relates classes in $\P^m$, 
$\P^n$ to classes in $\P^m\times \P^n$, while the polynomial $G(T)$
refers to a class in $\P^{m+n}$. While $\P^m\times \P^n$
and $\P^{m+n}$ can be related in a straightforward way by blow-ups
and blow-downs, tracking the fate of CSM classes across blow-up
operations is in itself a nontrivial (and worthy) task. One might 
optimistically think
that if a locally closed set avoids the center of a blow-up, then the
CSM class of its preimage ought to be the pull-back of its CSM class;
this is not true in general, as simple examples show. The fact that it
{\em is\/} true in certain cases is what we prove in \cite{AlMa},
Corollary~4.4, and this result is crucial for the computation of CSM 
classes of graph hypersurfaces of banana graphs. We know of no
general result of the same type handling the present situation, so we
have to provide a rather ad-hoc argument to prove the formula in
Lemma~\ref{lemmahomo}. Kwieci\'nski's product formula will be
an ingredient in our proof.

By Lemma~\ref{lemmahomo}, we are reduced to dealing with affine
cones over (nonempty) projective varieties $X\subseteq \P^{m-1}$. 
We begin by relating $G_{\hat X}(T)$ to the CSM class of $X$.

\begin{lem}\label{rewrilem}
Let $X\subseteq \P^{m-1}$ be a nonempty subvariety, and let $f(h)$ be
the polynomial of degree~$<m$ in the hyperplane class $h$ of $\P^{m-1}$, 
such that
\[
\csm(\one_X)= f(h)\cap [\P^{m-1}]\quad.
\]
Then
\[
h^m G_{\hat X}(1/h) = f(h) + h^m\quad.
\]
\end{lem}

\begin{proof}
Consider the cone $C(X)$ of $X\subseteq \P^{m-1}$ in $\P^m$.
By Proposition~5.2 in \cite{AlMa},
\[
\csm(\one_{C(X)}) = ((1+h) f(h)+h^m)\cap [\P^m]\quad,
\]
where $h$ denotes the hyperplane class in $\P^m$. Since $\hat X
\subseteq \A^m$ may be realized as the complement of $X$ in 
$C(X)$, 
\[
\csm(\one_{\hat X})= ((1+h) f(h) + h^m-h f(h))\cap [\P^m]
=(f(h)+h^m)\cap [\P^m]\quad.
\]
Since $h^k\cap [\P^{m-k}]$ corresponds to $T^{m-k}$ in 
Definition~\ref{Gdef}, 
\[
f(h)+h^m = h^m G_{\hat X}(1/h)\quad,
\]
as stated.
\end{proof}

Next, we relate the affine product of varieties to the projective {\em join.\/}
View $\P^{m-1}$, $\P^{n-1}$ as disjoint subspaces of $\P^{m+n-1}$.
For $X\subseteq\P^{m-1}$, $Y\subseteq\P^{n-1}$, we will denote by 
$\jXY$ (the `join' of $X$ and $Y$) the union of all lines connecting 
points of $X$ to points of $Y$ in~$\P^{m+n-1}$.

\begin{lem}\label{joinprod}
The product $\hat X \times \hat Y\subseteq \A^{m+n}$ is the affine cone
over the join $J(X,Y)\subseteq \P^{m+n-1}$.
\end{lem}

\proof
Denote by $(x_1:\ldots:x_m:y_1:\ldots: y_n)$ the points of $\P^{m+n-1}$;
identify $\P^{m-1}$ with the set of points $(x_1:\ldots:x_m:0:\ldots: 0)$ and
$\P^{n-1}$ with the set of points $(0:\ldots:0:y_1:\ldots: y_n)$.
If $(x_1:\ldots:x_m)\in X$ and $(y_1:\ldots:y_n)\in Y$, the points of the
line in $\P^{m+n-1}$ joining these two points are all and only the points
\[
(sx_1:\ldots:sx_m:ty_1:\ldots:t y_n)
\]
as $(s:t)$ varies in $\P^1$. It follows that a point $(x_1:\ldots:x_m:y_1:\ldots: y_n)$ 
is a point of $J(X,Y)$ if and $(x_1:\ldots:x_m)$ satisfies the homogeneous
ideal of $X$ in $\P^{m-1}$ and $(y_1:\ldots:y_n)$ satisfies the homogeneous
ideal of $Y$ in $\P^{n-1}$. This is the case if and only if
\[
(x_1,\ldots,x_m, y_1,\ldots, y_n)\in \hat X\times \hat Y \subseteq \A^{m+n}\quad,
\]
and this shows that the affine cone over $J(X,Y)$ is $\hat X\times \hat Y$.
\endproof

By Lemma~\ref{joinprod}, the sought-for formula
in Lemma~\ref{lemmahomo} may be rewritten as
\[
G_{\widehat\jXY}(T)=G_{\hat X}(T)\cdot G_{\hat Y}(T)\quad;
\]
or, equivalently (after a change of variable and harmless manipulations):
\begin{equation}\label{convtoH}
H^{m+n} G_{\widehat\jXY}(1/H)-H^{m+n}=
H^m G_{\hat X}(1/H)\cdot H^n G_{\hat Y}(1/H)-H^{m+n}
\end{equation}
for all nonempty $X\subseteq\P^{m-1}$, $Y\subseteq \P^{n-1}$. Here $H$
is simply a variable; but the two sides of the identity are polynomials of
degree $< (m+n)$ in $H$, so formula~\eqref{convtoH} may be 
verified by interpreting~$H$ as the hyperplane class in $\P^{m+n-1}$.
This formulation and Lemma~\ref{rewrilem}
reduce the proof of Theorem~\ref{mainhomo} to the
following computation of the CSM class of a join:

\begin{thm}\label{thmjoin}
Let $\P^{m-1}$, $\P^{n-1}$ be disjoint subspaces of $\P^{m+n-1}$,
and let $X\subseteq \P^{m-1}$, $Y\subseteq \P^{n-1}$ be nonempty
subvarieties. Let $f(H)$, resp.~$g(H)$ be polynomials such that 
\[
\csm(\one_X)=H^n f(H) \cap [\P^{m+n-1}]\quad, \quad
\csm(\one_Y)=H^m g(H) \cap [\P^{m+n-1}]\quad.
\]
Then
\[
\csm(\one_{\jXY}) = \left((f(H)+H^m)(g(H)+H^n)-H^{m+n}\right)
\cap [\P^{m+n-1}]\quad.
\]
\end{thm}

This is a result of independent interest, and its proof is deferred to \S\ref{joinpf}.
As argued in this section, Theorem~\ref{thmjoin} establishes 
Theorem~\ref{mainhomo}, concluding the proof that $C_\Gamma(T)$
satisfies the Feynman rules and the other properties listed in this section.

\section{Renormalization for algebro-geometric Feynman rules}\label{RenSec}

The Connes--Kreimer theory \cite{CoKr} (see also a detailed account in \S 1 of \cite{CoMa-book}) 
shows that the BPHZ procedure of renormalization of dimensionally regularized Feynman integrals
can be formulated as a Birkhoff factorization in the affine group scheme dual to the Connes--Kreimer
Hopf algebra of Feynman graphs. The explicit recursive formula for the Birkhoff factorization proved
by Connes and Kreimer in \cite{CoKr} gives a multiplicative splitting of an algebra homomorphism
$U: \cH \to \cK$, with $\cK$ the field of convergent Laurent series, as
\begin{equation}\label{Birkhoff}
 U = (U_- \circ S) \star U_+ 
\end{equation}
where $S$ is the antipode in the Connes--Kreimer Hopf algebra $\cH$ and $U_\pm : \cH \to \cK_\pm$
are algebra homomorphisms with values, respectively, in the algebra of convergent power series 
$\cK_+$ and the polynomial algebra $\cK_-=\C[z^{-1}]$. The product $\star$ is dual to the coproduct 
$\Delta$ of $\cH$ by $(U_1\star U_2)(X)=(U_1\otimes U_2)(\Delta(X))$.

The proof that the $U_\pm$, given in 
\cite{CoKr} by a recursive formula, are algebra homomorphisms uses the Rota--Baxter identity satisfied by the operator of projection of a Laurent series onto its polar part. The argument of Connes--Kreimer can therefore be easily generalized, in essentially the same form (see \cite{EFGK}), to the case
of algebra homomorphisms $U: \cH \to \cR$, with $\cH$ a polynomial ring in the 1PI graphs and
$\cR$ a Rota--Baxter ring of weight $-1$. We recall briefly how the renormalization procedure 
works.

A Rota--Baxter ring of weight $\lambda$ is a commutative ring $\cR$ endowed with 
a linear operator $\fT: \cR \to \cR$ satisfying the Rota--Baxter 
identity 
\begin{equation}\label{RBop}
\fT(x)\fT(y) = \fT(x \fT(y)) + \fT(\fT(x) y) + \lambda \fT(xy).
\end{equation}
Such an operator is called a Rota--Baxter operator of weight $\lambda$. 

Let $\cH$ denote the polynomial ring generated over $\Z$ by
the 1PI graphs, endowed with the coproduct 
\begin{equation}\label{coprod}
\Delta(\Gamma)= \Gamma\otimes 1 + 1 \otimes \Gamma + \sum_{\gamma\subset \Gamma} \gamma \otimes \Gamma/\gamma.
\end{equation}
Here the proper subgraphs $\gamma \subset \Gamma$ are possibly multiconnected, with
components that are 1PI. This is just slightly different from the Connes--Kreimer coproduct in
as we are not fixing a Lagrangian for a scalar field theory, hence we do not restrict only to
subgraphs such that $\Gamma/\gamma$ is still a Feynman graph of the given theory. In this
sense, it is similar to the Hopf algebras of graphs considered in \cite{JoRo} 
\cite{Rota}. The ring $\cH=\oplus_{n\geq 0} \cH_n$ is graded by the number $n=\# E_{int}(\Gamma)$ 
of internal edges of the graph and the antipode is defined inductively as
$$ S(\Gamma)= - \Gamma - \sum_{\gamma\subset \Gamma} S(\gamma) \,\, \Gamma/\gamma. $$

We then have the following result of Connes--Kreimer \cite{CoKr} (see also \cite{EFGK} for the
formulation in Rota--Baxter terms). 

\begin{prop}\label{renorm}
Suppose given a ring homomorphism $U: \cH \to \cR$, with $\cH$ as above and $\cR$ a Rota--Baxter ring of weight $-1$. Let $\cR_-$ denote the ring obtained by adjoining a unit to the ring $\fT \cR$
and let $\cR_+$ be the ring $\cR_+= (1-\fT)\cR$. Then the recursive formulae
\begin{equation}\label{BPHZ1}
U_-(\Gamma) = - \fT \left( U(\Gamma) + \sum_{\gamma \subset \Gamma} U_-(\gamma) U(\Gamma/\gamma)  \right)
\end{equation}
\begin{equation}\label{BPHZ2}
U_+(\Gamma) = (1 -\fT) 
\left( U(\Gamma) + \sum_{\gamma \subset \Gamma} U_-(\gamma) U(\Gamma/\gamma)  \right)
\end{equation}
determine a Birkhoff factorization \eqref{Birkhoff} into algebra homomorphisms 
$U_\pm : \cH\to \cR_\pm$. There is a unique such factorization
satisfying the normalization condition $\epsilon_- \circ U_- = \epsilon$, where 
$\epsilon_- :\cR_- \to \Z$ is the augmentation and $\epsilon$ is the counit of $\cH$,
defined by $\epsilon(X)=0$ for $\deg(X)>0$.
\end{prop}

In the case of the dimensionally regularized Feynman integrals, the $U_-$ gives
the counterterms and the evaluation of the convergent power series $U_+(\Gamma)$,
\begin{equation}\label{evUplus}
U_+(\Gamma) |_{z=0},
\end{equation}
gives the renormalized value of the Feynman integral $U(\Gamma)$. 

We can apply the same procedure to the algebro--geometric Feynman rules, using suitable
Rota--Baxter operators on the target ring. This will give new multiplicative invariants of graphs
obtained by following the same BPHZ procedure that governs the renormalization of divergent
Feynman integrals.

\smallskip

For example, consider the motivic Feynman rule $\bU(\Gamma)=[\A^n\smallsetminus \hat X_\Gamma] \, \bL^{-n}$ in $K_0(\cV_\C)[\bL^{-1}]$. In the ring $K_0(\cV_\C)[\bL^{-1}]$ we can still consider the Rota--Baxter operator of projection onto the polar part (in the variable $\bL$). The renormalized Feynman
rule $\bU_+(\Gamma)$ defined as in \eqref{BPHZ2} defines a class in $K_0(\cV_\C)$ and the ``renormalized value of the Feynman integral" \eqref{evUplus} defines a class in $\Z[SB]$, 
\begin{equation}\label{UplusSB}
\bU_+(\Gamma) |_{\bL =0} = (1 -\fT) 
\left( U(\Gamma) + \sum_{\gamma \subset \Gamma} U_-(\gamma) U(\Gamma/\gamma)  \right) |_{\bL =0} 
\in \Z[SB]= K_0(\cV_\C)/(\bL). 
\end{equation}
Notice that the parts of $[\A^n \smallsetminus \hat X_\Gamma]$, 
$[\A^{n_1}\smallsetminus \hat X_\gamma]$ and
$[\A^{n_2}\smallsetminus \hat X_{\Gamma/\gamma}]$  that are contained in the ideal 
$(\bL)\subset K_0(\cV_0)$ contribute cancellations to the $\bL^n$ in the denominator.
It is possible that this invariant and the Birkhoff factorization of $\bU(\Gamma)$ may help 
to detect the presence of non-mixed-Tate strata in the graph hypersurface $X_\Gamma$ 
coming from the contributions of hypersurfaces of smaller graphs $\gamma \subset \Gamma$ 
or quotient graphs $\Gamma/\gamma$, 

\smallskip

For invariants like $C_\Gamma(T)$ that take values in polynomial rings, one can
consider different kinds of Rota--Baxter operators. For example, consider the basis
of $\Q[T]$ as a $\Q$-vector space, given by the polynomials
$$ \pi_n(T) = \frac{T(T+1)\cdots (T+n-1)}{n!}, \ \ \forall n\geq 1, \ \ \ \pi_0(T)=1. $$
The linear operator $\fT (\pi_n) = \pi_{n+1}$ is a non-trivial Rota--Baxter operator of weight 
$-1$ on the polynomial ring $\Q[T]$ (see \cite{Guo}). One can then apply the BPHZ procedure
with respect to this or other interesting Rota--Baxter operators to have a Birkhoff factorization
of the given invariant with respect to an assigned Rota--Baxter structure. We do not pursue
further in this paper the meaning of BPHZ with respect to different possible Rota--Baxter 
operators, but we only remark that algebro--geometric Feynman rules provide a supply of
examples where one can abstractly study the properties of the BPHZ renormalization 
procedure. For example, the question of whether the Grothendieck ring of varieties
$K_0(\cV_k)$ or our Grothendieck ring of immersed conical varieties $\cF$ admit a Rota--Baxter
structure of weight $-1$ appears to be a problem of independent interest.

\smallskip

Finally, we can consider again the possible definition \eqref{UGammaPer} of Feynman
rules with values in the algebra $\cP$ of periods and the problem of the divergences
caused by the nontrivial intersections of the domain of integration $\sigma_n$ with the
hypersurface $\hat X_\Gamma$. In \cite{Mar} a regularization for Feynman integrals
of the form \eqref{FeyParp0} was proposed based on replacing the part of the integral
that takes place in a neighborhood of the hypersurface $\hat X_\Gamma$ of the form
$D_\epsilon(\hat X_\Gamma) =\cup_{s\in \Delta_\epsilon^*} \hat X_\Gamma(s)$, 
given by the level sets $\hat X_\Gamma(s)=\{ t\in \A^n \,|\, \Psi_\Gamma(t)=s \}$
for $s\in \Delta_\epsilon^*$ a small punctured disk of radius $\epsilon >0$, with an
integral on a Leray coboundary $\cL_\epsilon(\sigma_n)=\pi^{-1}(\sigma_n \cap X_\epsilon)$,
for $\pi_\epsilon: \partial D_\epsilon(\hat X_\Gamma) \to \hat X_\Gamma(\epsilon)$ the
circle bundle projection. This has the effect of replacing the (divergent) integration on
the locus $\sigma_n \cap \hat X_\Gamma$ with one on circles around the singular locus.
By the results of \cite{Arn} Part III, \S 10.2 and Theorem 4.4 of \cite{Mar},
the resulting integral $\bU(\Gamma)(\epsilon)$ extends to a meromorphic function of 
$\epsilon$ in a small neighborhood of $\epsilon=0$,
with a pole at $\epsilon=0$. One can then apply the BPHZ renormalization procedure,
with $\fT$ the projection onto the polar part of the Laurent series in $\epsilon$ and
obtain a renormalized $\bU_+(\Gamma)$.

\section{The partition function and Tate motives}\label{ZSec}

In quantum field theory it is customary to consider the full partition function of the theory,
arranged in an asymptotic series by loop number or another suitable grading of the
Hopf algebra of Feynman graphs, instead of looking only at  the contribution of individual 
Feynman graphs. Besides the loop number $\ell=b_1(\Gamma)$, other suitable gradings 
$\delta(\Gamma)$ are given by the number $n=\# E_{int}(\Gamma)$ of internal edges, 
or by $\# E_{int}(\Gamma) - b_1(\Gamma)=\# V(\Gamma)-b_0(\Gamma)$, 
the number of vertices minus the number of connected components  
(\cf~\cite{CoMa-book} p.77).

When one considers motivic Feynman rules, these partition functions appear to be
interesting analogs of the motivic zeta functions considered in \cite{Kap}, \cite{LaLu}.
For instance, one can consider a partition function given by the formal series
\begin{equation}\label{ZTmot}
Z(t) = \sum_{N\geq 0} \sum_{\delta(\Gamma)=N} \frac{\bU(\Gamma)}{\# \Aut(\Gamma)}\, t^N ,
\end{equation}
where $\delta(\Gamma)$ is any one of the gradings on the Hopf algebra of Feynman graphs
described above and where $\bU(\Gamma)=[\A^n\smallsetminus \hat X_\Gamma] \in K_0(\cV_k)$.
Given a motivic measure $\mu: K_0(\cV_k) \to \cR$, this gives a zeta function with values in 
$\cR[[t]]$ of the form
$$ Z_{\cR}(t) = \sum_{N\geq 0} \sum_{\delta(\Gamma)=N} \frac{\mu(\bU(\Gamma))}{\# \Aut(\Gamma)}\, t^N. $$

Of particular interest, in view of the recent results of \cite{Blo2}, is the case where one restricts
the class of graphs to connected graphs without looping edges and without multiple edges
and takes the grading $\delta(\Gamma)=\# V(\Gamma)-b_0(\Gamma)$. In this case, one is
considering the zeta function
\begin{equation}\label{ZmotSN}
Z(t) = \sum_{N\geq 1} \frac{t^N}{N!} \sum_{\#V(\Gamma)=N} \bU(\Gamma) \frac{N!}{\# \Aut(\Gamma)}.
\end{equation}
The result of \cite{Blo2} shows that 
\begin{equation}\label{SumGraphsTate}
S_N = \sum_{\#V(\Gamma)=N} [X_\Gamma] \frac{N!}{\# \Aut(\Gamma)}
\end{equation}
is in the Tate part of the Grothendieck ring, $S_N \in \Z [\bL]$. It then follows that
the zeta function $Z(t)$ above takes values in $\Z[\bL][[t]]$. 

One can investigate the behaviour of these ``motivic zeta functions" by the same techniques
used in \cite{LaLu} to study the original motivic zeta function defined by Kapranov 
in~\cite{Kap}.

\section{The formula for CSM classes of joins}\label{joinpf}

This section is devoted to the proof of Theorem~\ref{thmjoin}.
We first recall the statement. 

Let $X\subseteq \P^{m-1}$, $Y\subseteq \P^{n-1}$ be 
nonempty subvarieties, and view
$\P^{m-1}$, $\P^{n-1}$ as disjoint subspaces of $\P^{m+n-1}$. The
task is to compute the push-forward to $\P^{m+n-1}$ of the
Chern-Schwartz-MacPherson class of the join $\jXY$, defined
as the union of the lines in $\P^{m+n-1}$ connecting all points of $X$ to
all points of $Y$. The class will be expressed as a polynomial in the
class $H$ of a hyperplane in $\P^{m+m-1}$, obtained in terms
of the polynomials similarly giving the Chern-Schwartz-MacPherson
classes of $X$ in $\P^{m-1}$, $Y$ in $\P^{n-1}$.

We will denote by $h$ the hyperplane class in $\P^{m-1}$, and by $k$
the hyperplane class in~$\P^{n-1}$. Let $f(h)$, $g(k)$ be polynomials
of degree $<m$, resp.~$<n$, such that
\begin{align*}
\csm(\one_X)= f(h)\cap [\P^{m-1}] \quad, \\
\csm(\one_Y)= g(k)\cap [\P^{n-1}] \quad.
\end{align*}

Theorem~\ref{thmjoin} states the following result:
\begin{equation}\label{joinformula}
\csm(\one_\jXY) = \left((f(H)+H^m)(g(H)+H^n)-H^{m+n}\right)
\cap [\P^{m+n-1}]\quad.
\end{equation}
The rest of this section is devoted to the proof of this formula.

\begin{ex}\label{coneEx} {\rm
If $Y=\P^{n-1}$, then $\jXY$ is the cone $C^n(X)$ on $X$, with vertex 
$\P^{n-1}$. Since $c(T\P^{n-1})=(1+H)^n-H^n$, 
\eqref{joinformula} gives
\[
\csm(C^n(X))=\left((1+H)^n (f(H)+H^m)-H^{m+n}\right)\cap [\P^{m+n-1}]\quad.
\]
where a push-forward is understood. In particular, for $n-1=0$ (so $C^1(X)
=C(X)$
is just an `ordinary' cone in projective space) this agrees with the formula for
cones given in Proposition~5.2 of~\cite{AlMa}.}
\end{ex}

\begin{ex}\label{Pasjoin}{\rm
For $X=\P^{m-1}$, $Y=\P^{n-1}$, the join $\jXY$ is the whole
of $\P^{m+n-1}$. Theorem~\ref{thmjoin} gives
\[
\csm(\P^{m+n-1})=\left((1+H)^m(1+H)^n-H^{m+n}\right) \cap[\P^{m+n-1}]\quad,
\]
as it should.}
\end{ex}

We will realize the join of $X$ and $Y$ as a projection of a $\P^1$-bundle 
over $X\times Y$. Consider the rational map
\[
\P^{m+n-1} \dashrightarrow \P^{m-1}\times \P^{n-1}
\]
given by
\[
(x_1: \ldots : x_m :  y_1:  \ldots :  y_n)
\mapsto 
((x_1: \ldots : x_m), (y_1:  \ldots :  y_n))\quad;
\]
this is well-defined away from the union $\P^{m-1}\cup \P^{n-1}$
consisting of points where either
\[
y_1=\cdots = y_n=0\quad
\]
or
\[
x_1=\cdots = x_m=0\quad.
\]
Letting $B\ell$ be the blow-up of $\P^{m+n-1}$ along these two
linear subspaces, we obtain a diagram
\[
\xymatrix@C=10pt{
& B\ell \ar[dl]_\pi \ar[dr]^\rho \\
\P^{m+n-1} \ar@{-->}[rr] & & \P^{m-1}\times \P^{n-1}
}
\]
resolving the given rational map, and realizing $B\ell$ as a $\P^1$-bundle
over $\P^{m-1}\times \P^{n-1}$. Concretely, $\rho^{-1}(p,q)$ may be
identified with the (proper transform of the) line in $\P^{m+n-1}$
joining $p\in \P^{m-1}$ to $q\in \P^{n-1}$.

Summary of the argument: we will use Kwieci\'nski's product formula 
(\cite{Kwi}) and Yokura's Riemann-Roch for Chern-Schwartz-MacPherson 
classes (\cite{Yok}) to compute the class of the inverse image of $X\times Y$
in $B\ell$. The formula for the class of $\jXY$ will follow from this and the
basic functoriality property of CSM classes.
\smallskip

We first collect the necessary ingredients. 

As noted above, $h$ and $k$
denote respectively the hyperplane classes in $\P^{m-1}$, $\P^{n-1}$;
we use the same letters to denote their pull-backs to the product
$\P^{m-1}\times \P^{n-1}$, and to $B\ell$.

\begin{lem}\label{Kwiapp}
With notation as above, 
\[
\csm(\one_{X\times Y})=f(h) g(k) \cap [\P^{m-1}\times \P^{n-1}]\quad.
\]
\end{lem}

\proof
There is a natural map $A_* X \otimes A_* Y \to A_*(X\times Y)$ 
(\S1.10 in \cite{Ful}). By Kwieci\'nski's theorem in \cite{Kwi} 
(cf.~also Theorem~4.1 in \cite{Alu1}), this map sends $\csm(X)\otimes 
\csm(Y)$ to $\csm(X\times Y)$. Pushing forward to the ambient
product of projective spaces, this says that $\csm(\one_{X\times Y})$
is the image of $(f(h)\cap [\P^{m-1}])\otimes (g(k)\cap [\P^{n-1}])$;
this is the statement.
\endproof

Viewing $B\ell$ as the blow-up of $\P^{m+n-1}$ along the skew 
$\P^{m-1}$ and $\P^{n-1}$, let $E$ be the component of the exceptional 
divisor over $\P^{m-1}$, and $F$ the component over $\P^{n-1}$.
Denote by $H$ the hyperplane class in $\P^{m+n-1}$, as well as
its pull-back to $B\ell$. 

The classes $H$, $h$, $k$, $E$, $F$ in $B\ell$ are not independent:

\begin{lem}\label{relblup}
$h=H-F$, and $k=H-E$.
\end{lem}

\proof
The projection $\P^{m+n-1} \dashrightarrow \P^{m-1}$ is resolved
by the blow-up $\widetilde \P^{m+n-1}$ of $\P^{m+n-1}$ along $\P^{n-1}$. 
It is clear that (the pull-back of) $h$ agrees with $H-F$ in this blow-up, 
where $F$ denotes the exceptional divisor over $\P^{n-1}$. This relation 
pulls back to the same relation in $B\ell$, which may be realized as
the pull-back of $\widetilde \P^{m+n-1}$ along the inverse image of
$\P^{m-1}$.

This proves the first relation. The second relation is obtained similarly.
\endproof

By Lemma~\ref{Kwiapp} and~\ref{relblup}, the pull-back of 
$\csm(\one_{X\times Y})$ to $B\ell$ is given by
\[
f(H-F) g(H-E) \cap [B\ell]\quad.
\]
The CSM class of $\rho^{-1}(X\times Y)$ may be obtained from
this by applying a result of Shoji Yokura. For this, we note that
$B\ell$ is smooth over $\P^{m-1}\times \P^{n-1}$, and more precisely
$B\ell$ may be realized as the projectivization of $\cO(-h)\oplus \cO(-k)$.
With this choice, the pull-back of $\cO(H)$ agrees with the tautological
bundle $\cO(1)$ on $B\ell \cong \P(\cO(-h)\oplus \cO(-k))$.

\begin{lem}\label{YokRR}
\begin{equation}\label{inBl}
\csm(\one_{\rho^{-1}(X\times Y)})=(1+F)f(H-F) (1+E)g(H-E)\cap [B\ell]\quad.
\end{equation}
\end{lem}

\proof
Write $\cE=\cO(-h)\oplus \cO(-k)$, so $B\ell\cong \P(\cE)$.
By Theorem~2.2 in \cite{Yok}, CSM classes behave like ordinary
Chern classes through smooth morphisms: thus,
\[
\csm(\one_{\rho^{-1}(X\times Y)})=c(T_{B\ell\mid (\P^{m-1}\times \P^{n-1})})
\cap \rho^*(\csm(\one_{X\times Y}))\quad.
\]
The pull-back $\rho^*(\csm(\one_{X\times Y}))=f(H-F)g(H-E)\cap [B\ell]$ was 
computed above. The relative tangent bundle 
$T_{B\ell\mid (\P^{m-1}\times \P^{n-1})}$ is computed by means of the 
Euler exact sequence (cf.~\cite{Ful},~B.5.8)
\[
\xymatrix{
0 \ar[r] & \cO \ar[r] & \rho^*\cE\otimes \cO(H) \ar[r] &
T_{B\ell\mid (\P^{m-1}\times \P^{n-1})} \ar[r] & 0
}
\]
and gives (as $\cE=\cO(-h)\oplus \cO(-k)$)
\[
c(T_{B\ell\mid (\P^{m-1}\times \P^{n-1})})=c(\rho^* \cE\otimes \cO(H))
=(1-h+H)(1-k+H)\quad.
\]
The statement follows from this and Lemma~\ref{relblup}.
\endproof

\begin{ex}{\rm
For $X=\P^{m-1}$, $Y=\P^{n-1}$, we have $\rho^{-1}(X\times Y)=B\ell$,
and $f(h)=(1+h)^m-h^m$, $g(k)=(1+k)^n-k^n$. Noting that $h^m=0$,
$k^n=0$, formula~\eqref{inBl} gives
\[
c(TB\ell)\cap [B\ell]=(1+F)(1+H-F)^m (1+E)(1+H-E)^n\cap [B\ell]\quad. 
\]
This may also be obtained by two applications of Lemma~1.3
in \cite{Alu2}, since $\P^{m-1}$ and $\P^{n-1}$ are disjoint complete 
intersections in $\P^{m+n-1}$.
}
\end{ex}

These preliminary considerations yield the following statement.

\begin{lem}\label{summm}
Let $\pi:B\ell\to \P^{m+n-1}$ be the blow-up along two disjoint centers
$\P^{m-1}$, $\P^{n-1}$; let $E$, resp.~$F$ be the exceptional
divisors over these two centers; and let $H$ denote the hyperplane class
in $\P^{m+n-1}$, as well as its pull-back to $B\ell$. For $X\subseteq
\P^{m-1}$, $Y\subseteq \P^{n-1}$ nonempty subvarieties, let $f(H)$, 
resp.~$g(H)$ be polynomial expressions of degrees~$<m$, 
resp.~$<n$ in $H$, such that
\[
\csm(\one_X)=f(H)\cap [\P^{m-1}]\quad,\quad 
\csm(\one_Y)=g(H)\cap [\P^{n-1}]
\]
as classes in $\P^{m+n-1}$.
Finally, let $\jXY\hookrightarrow \P^{m+n-1}$ be the join of $X$ 
and $Y$ in~$\P^{m+n-1}$. Then
\begin{multline*}
\csm(\one_\jXY)=\pi_*\left(
(1+F)(1+E)f(H-F)g(H-E)\cap [B\ell]
\right)\\
- (\chi(Y)-1) f(H)\cap [\P^{m-1}] 
- (\chi(X)-1) f(H)\cap [\P^{n-1}] 
\quad.
\end{multline*}
\end{lem}

In this statement, $\chi(X)$ and $\chi(Y)$ denote the Euler characteristics
of $X$ and $Y$, respectively.

\begin{proof}
Realize the join $\jXY$ as the image of $\rho^{-1}(X\times Y)$ in 
$\P^{m+n-1}$. Denote by $\overline\pi: \rho^{-1}(X\times Y)
\to \jXY$ the restriction of $\pi$. Then $\overline \pi$ is proper and 
birational, and contracts
\[
E\cap \rho^{-1}(X\times Y)\quad\text{to}\quad X\subseteq \P^{m-1}\quad,
\]
\[
F\cap \rho^{-1}(X\times Y)\quad\text{to}\quad Y\subseteq \P^{n-1}\quad.
\]
Now, $E\cap \rho^{-1}(X\times Y)\cong X\times Y$, and the contraction
corresponds to the projection $X\times Y \to X$. Similarly, the contraction
$F\cap \rho^{-1}(X\times Y)\quad\text{to}\quad Y$ corresponds to the
projection $X\times Y\to Y$. Therefore, the fibers of 
$\overline\pi$ may be described as
follows:
\begin{align*}
p\not\in X\cup Y & \implies \overline\pi^{-1}(p)=\text{ a point}\\
p\in X & \implies \overline\pi^{-1}(p)\cong Y\\
p\in Y & \implies \overline\pi^{-1}(p)\cong X
\end{align*}
In terms of constructible functions, this says
\begin{align*}
\overline\pi_*(\one_{\rho^{-1}(X\times Y)})
&=\one_{\jXY\smallsetminus (X\cup Y)}+\chi(Y) \one_X+\chi(X) \one_Y\\
&=\one_{\jXY}+(\chi(Y)-1) \one_X+(\chi(X)-1) \one_Y\quad.
\end{align*}
By the functoriality property of Chern-Schwartz-MacPherson's classes, it follows
that
\[
\overline\pi_* \csm(\one_{\rho^{-1}(X\times Y)})
=\csm(\jXY)+(\chi(Y)-1)\csm(X) + (\chi(X)-1) \csm(Y)\quad.
\]
The statement follows immediately from this, together with 
Lemma~\ref{YokRR}.
\end{proof}

The challenge now is to evaluate the push-forward
\[
\pi_*((1+F)(1+E) f(H-F) g(H-E)\cap [B\ell])\quad.
\]
Since $f$ and $g$ are polynomials, this is a sum of terms
\[
\pi_*((1+F)(1+E) (H-F)^i (H-E)^j\cap [B\ell])\quad.
\]
This push-forward can be executed in two steps, since $\pi$ may be
viewed as a composition $\pi=\pi_2\circ \pi_1$ of the blow-up $\pi_1$
of $\P^{m+n-1}$ along $\P^{m-1}$, followed by the blow-up
$\pi_2$ of the resulting variety along (a locus isomorphic to)
$\P^{n-1}$. Both steps match the following template:

\begin{lem}\label{pfblup}
Let $p: \widetilde V \to V$ be the blow-up of a scheme $V$ along a subscheme
$W$ of codimension~$r$. Assume $W$ has class $H^r$, where $H$
is a divisor class in $V$. Denote by the same letter $H$ the pull-back
of this divisor class to $\widetilde V$, and let $D$ be the exceptional divisor. 
Then
\[
p_*((H-D)^j)=\left\{
\aligned
H^j\quad & 0\le j < r \\
0\quad & j\ge r
\endaligned
\right.
\quad,\quad
p_*(D(H-D)^j)=\left\{
\aligned
H^r\quad & j = r-1 \\
0\quad & j\ne r-1
\endaligned
\right.\quad.
\]
\end{lem}

\begin{proof}
By the birational invariance of Segre classes (Proposition~4.2(a) in \cite{Ful}), 
\[
p_*\left( \frac{D}{1+D}\cap [\widetilde V]\right) = 
s(W,V) = \frac{H^r}{(1+H)^r}\cap [V]\quad, 
\]
and hence
\[
p_* \left(\frac{1}{1+D}\cap [\widetilde V]\right) = 
\left(1-\frac{H^r}{(1+H)^r}\right)\cap [V]\quad.
\]
Introducing a bookkeeping variable $v$, we have
\[
p_* \left(\frac{1}{1+vD}\cap [\widetilde V]\right)= 
\left(1-\frac{(vH)^r}{(1+vH)^r}\right)\cap [V]\quad:
\]
indeed, multiplying $D$ by $v$ on the left has the effect of multiplying
every term of codimension~$j$ by $v^j$, and this is the same effect
obtained by multiplying $H$ by $v$ on the right. By the projection
formula, $v$ may be replaced by any expression in $H$ on the right
and by its pull-back on the left, still yielding a correct identity.
Apply this observation to
\[
\sum_{j\ge 0} (H-D)^j = \frac{1}{1+D-H}=\frac{\frac{1}{1-H}}{1+\frac{D}{1-H}}
\quad,
\]
with $v=\frac 1{1-H}$:
\begin{align*}
p_*\left( \sum_{j\ge 0} (H-D)^j \cap[\widetilde V]\right) &= \frac{1} {1-H}\cdot
p_* \left(\frac{1}{1+\frac 1{1-H}D}\cap [\widetilde V]\right) \\
&= \frac{1} {1-H}\cdot\left(1-\frac{(\frac 1{1-H}H)^r}
{(1+\frac 1{1-H}H)^r}\right)\cap [V]\\
&=\frac{1} {1-H}\cdot (1-H^r)\cap [V]\\
&=(1+H+\cdots +H^{r-1})\cap [V]\quad.
\end{align*}
This establishes the first formula. The second formula follows 
immediately from this, by observing that
\[
D(H-D)^j = H (H-D)^j - (H-D)^{j+1}\quad.
\]
\end{proof}

Returning to our analysis of intersections in $B\ell$, Lemma~\ref{pfblup} gives

\begin{lem}\label{contract}
\[
\pi_*\left((H-F)^i (H-E)^j \cap [B\ell]\right)
=\left\{
\aligned
H^{i+j}\cap [\P^{m+n-1}]\quad & \text{if $0\le i< m$ and $0\le j< n$} \\
0 \qquad\qquad\quad & \text{otherwise}
\endaligned
\right.
\]
\[
\pi_*\left(E(H-F)^i (H-E)^j \cap [B\ell]\right)
=\left\{
\aligned
H^{i+n}\cap [\P^{m+n-1}]\quad & \text{if $0\le i< m$ and $j=n-1$} \\
0 \qquad\qquad\quad & \text{otherwise}
\endaligned
\right.
\]
\[
\pi_*\left(F(H-F)^i (H-E)^j \cap [B\ell]\right)
=\left\{
\aligned
H^{j+m}\cap [\P^{m+n-1}]\quad & \text{if $0\le j< n$ and $i=m-1$} \\
0 \qquad\qquad\quad & \text{otherwise}
\endaligned
\right.
\]
\[
\pi_*\left(EF(H-F)^i (H-E)^j \cap [B\ell]\right)
=0.
\]
\end{lem}

\begin{proof}
The last formula follows from the fact that $EF=0$ (the two exceptional
divisors are disjoint). The others are each obtained by applying 
Lemma~\ref{pfblup} twice. For example, note that 
\[
\pi_*\left((H-F)^i (H-E)^j \cap [B\ell]\right)
=\pi_{1*}\left((H-E)^j \cdot \pi_{2*} \left((H-F)^i \cap [B\ell]\right)
\right)
\]
by the projection formula, since $H$, $E$ are pull-backs from the first 
blow-up. Hence, Lemma~\ref{pfblup} evaluates this class to
\[
\pi_{1*}\left((H-E)^j \cdot H^i\right)
\]
if $0\le i< m$ ($m=$ the codimension of $\P^{n-1}$) and $0$ 
otherwise; and another application of Lemma~\ref{pfblup} evaluates
this to $H^{i+j}$ if both $0\le i< m$ and $0\le j< n$, and $0$
otherwise. The remaining two formulas are handled similarly.
\end{proof}

We are finally ready to prove Theorem~\ref{thmjoin}.

\begin{proof}[Proof of Theorem~\ref{thmjoin}]
We have to evaluate
\[
\pi_*\left(
(1+F)(1+E)f(H-F)g(H-E)\cap [B\ell]
\right)\quad.
\]
Let $f(x)=\sum_{i=0}^{m-1} a_i x^j$ and 
$g(x)=\sum_{j=0}^{n-1} b_j x^i$. Then
\begin{multline*}
\pi_*\left(
(1+F)(1+E)f(H-F)g(H-E)\cap [B\ell]
\right)\\
=\pi_*\left(
f(H-F)g(H-F)+E f(H-F)g(H-E) +F f(H-F)g(H-E) 
\right)
\end{multline*}
\begin{multline*}
=\sum_{i=0}^{m-1} \sum_{j=0}^{n-1} a_i b_j \pi_*\left(
(H-F)^i(H-E)^j\cap [B\ell]
\right)\\
+\sum_{i=0}^{m-1} \sum_{j=0}^{n-1} a_i b_j \pi_*\left(
E(H-F)^i(H-E)^j\cap [B\ell]
\right)\\
+\sum_{i=0}^{m-1} \sum_{j=0}^{n-1} a_i b_j \pi_*\left(
F(H-F)^i(H-E)^j\cap [B\ell]
\right)
\end{multline*}
\begin{multline*}
=\sum_{i=0}^{m-1} \sum_{j=0}^{n-1} a_i b_j H^{i+j}
+\sum_{i=0}^{m-1} a_i b_{n-1} H^{i+n}
+\sum_{j=0}^{n-1} a_{m-1} b_j H^{j+m}\\
=f(H)g(H)+\chi(Y) f(H)H^n+\chi(X) g(H)H^m\quad,
\end{multline*}
using Lemma~\ref{contract}, and the fact that 
$\chi(X)=\int \csm(X)=a_{m-1}$, 
$\chi(Y)=\int \csm(Y)=b_{n-1}$.
By Lemma~\ref{summm}, then,
\begin{align*}
\csm(\one_{\jXY})&=\left(f(H) g(H)+\chi(Y) f(H)H^n+\chi(X) g(H)H^m\right)
\cap [\P^{m+n-1}]\\
&\qquad \qquad - (\chi(Y)-1) f(H)\cap [\P^{m-1}] 
- (\chi(X)-1) f(H)\cap [\P^{n-1}]\\
&=\left(f(H)g(H)+ f(H)H^n+ g(H)H^m\right)
\cap [\P^{m+n-1}]\\
&=\left((f(H)+H^m)(g(H)+H^n)-H^{m+n}\right)\cap [\P^{m+n-1}]\quad.
\end{align*}
This establishes formula \eqref{joinformula}, and concludes the proof of 
Theorem~\ref{thmjoin}.
\end{proof}

\end{document}